\documentclass[lettersize,journal]{IEEEtran}
\usepackage[backend=bibtex,style=IEEEtran,natbib=true,style=numeric]{biblatex} 
\usepackage{graphicx} 
\usepackage{amsmath,amsfonts}
\usepackage{algorithmic}
\usepackage{algorithm}
\usepackage{array}
\usepackage{textcomp}
\usepackage{stfloats}
\usepackage{url}
\usepackage{verbatim}

\usepackage{multirow}
\usepackage{framed}
\usepackage{amssymb}
\usepackage{booktabs}
\usepackage{xcolor}
\usepackage{tikz}
\usepackage{adjustbox}
\usepackage{threeparttable}
\usepackage{orcidlink}
\newcommand{\mycircle}[1]{\tikz{\filldraw[draw=#1,fill=#1] (0,0) circle [radius=0.2em];}}

\newcommand{\mytriangle}[1]{\tikz{\filldraw[draw=#1,fill=#1] (0,0) -- (0.5em,0) -- (0.3em,0.5em);}}

\begin{document}

\title{Characterizing Data Visualization Literacy: a Systematic Literature Review}

\author{Sara Beschi~\orcidlink{0009-0006-4969-6565}, Davide Falessi~\orcidlink{0000-0001-8142-2253}, Silvia Golia~\orcidlink{0000-0003-0015-8126}, Angela Locoro~\orcidlink{0000-0002-6740-8620}
\thanks{Sara Beschi, Silvia Golia, and Angela Locoro are with the Department of Economics and Management, University of Brescia, \{sara.beschi,silvia.golia,angela.locoro\}@unibs.it, Davide Falessi is with the Department of Informatics Engineering, University of Rome Tor Vergata, davide.falessi@ing.uniroma2.it.}
\thanks{Manuscript received XXX, 2024; revised XXX}}

\markboth{Journal of \LaTeX\ Class Files,~Vol.~14, No.~8, August~2021}%
{Shell \MakeLowercase{\textit{et al.}}: A Sample Article Using IEEEtran.cls for IEEE Journals}

\maketitle

\IEEEpubid{0000--0000/00\$00.00~\copyright~2021 IEEE}

\begin{abstract}
With the advent of the data era, and of new, more intelligent interfaces for supporting decision making, there is a growing need to define, model and assess human ability and data visualizations usability for a better encoding and decoding of data patterns. Data Visualization Literacy (DVL) is the ability of encoding and decoding data into and from a visual language. Although this ability and its measurement are crucial for advancing human knowledge and decision capacity, they have seldom been investigated, let alone systematically. To address this gap, this paper presents a systematic literature review comprising 43 reports on DVL, analyzed using the PRISMA methodology. 
Our results include the identification of the purposes of DVL, its satellite aspects, the models proposed, and the assessments designed to evaluate the degree of DVL of people. 
Eventually, we devise many research directions including, among the most challenging, the definition of a (standard) unifying construct of DVL.

\end{abstract}


\section{Introduction}
\label{sec:intro}
\IEEEPARstart{W}{ith} the advent of the data era and of new, more intelligent interfaces for supporting decision-making, there is a growing need to define, model and assess human ability and data visualization usability for better encoding and decoding of data patterns.
The term \textit{Data Visualization Literacy} (DVL) stands for the ability to properly process information related to data visualizations (data viz from now on), i.e., encoding information into and decoding information from data viz~\cite{locoro2021visual}.
Since the DVL research area has been addressed in various contexts and by different research groups at different points in time, there are different terms for similar concepts, and this can lead to misunderstandings. For example, in this work we do not adhere to the definition of DVL provided by~\citet{Borner20191857} and~\citet{peppler2021cultivating}, i.e.,``the ability to make meaning from and interpret patterns, trends, and correlations in visual representations of data'', because it does not take into account the ability of \textit{encoding} visual information as well as \textit{decoding} it. 

Data viz encompasses many terms such as ``visualization'', ``chart'', ``graph'' and ``plot'', which subsumes all of the data viz artefacts that, from time to time, are found in the literature, e.g., diagrams, maps, bar charts, line charts, pie charts, and infographics containing them. Early references in the field of semiotics simply referred to them as \textit{graphics} (e.g., \cite{bertin1983semiology}), while avant-guarde designers, such as~\citet{tufte1990envisioning}, referred to them as \textit{data graphics}.  For the term \textit{Visual Literacy}, a standard exists\footnote{See, e.g., the Association of College and Research Libraries (ACRL) website, available at \url{www.ala.org/acrl/standards/visualliteracy.}}, which refers to the broader definition of ``abilities that enables an individual to effectively find, interpret, evaluate, use, and create images and visual media. Visual literacy skills equip a learner to understand and analyze the contextual, cultural, ethical, aesthetic, intellectual, and technical components involved in the production and use of visual materials. A visually literate individual is both a critical consumer of visual media and a competent contributor to a body of shared knowledge and culture''. A recent vibrant and important workshop~\cite{Ge2024Visualization} and many other studies ~\cite{camba2022identifying,firat2022interactive,firat2022vislite,lee2019correlation,cui2023adaptive} used the term \textit{Visualization Literacy} (VL) instead, which is associated with many targets, such as ``a person’s ability to do fundamental visualization tasks (e.g., retrieving values or making comparisons), correctly interpret data viz in the face of misinformation, or even create data viz.''. 
This difference in terminology may be grounded in that \textit{Visual Literacy} includes many kinds of visuals such as, for example, natural images (photography), visual arts, or schematization of natural phenomena for educational purposes (e.g., depicting the DNA helix, the main biochemistry processes, to show how they work, and the like). Many of these visuals are out of the scope of this paper and of those regarding VL. In particular, we intend to focus only on visuals related to the \textit{visual representations of data} as variables of any kind among nominal, ordinal, and discrete and continuous (interval and ratio) numbers, as in Steven's categorization~\cite{stevens1946theory}. Thus, this work is about DVL rather than \textit{Visual Literacy}, and we express our preference for the term \textit{Data Visualization Literacy} with respect to \textit{Visualization Literacy} as being less ambiguous. However, studies about \textit{Visual literacy} or VL cannot be discarded \textit{a priori}, as they include data viz or, better yet, they don't exclude them.

\IEEEpubidadjcol

In addition to defining DVL, we need an explicit and precise definition of its measurement. In this work, we adhere to the definition of DVL measurement provided by~\citet{mari2023measurement}: ``an empirical and informational process that is designed on purpose, whose input is an empirical property of an object (subject), and that produces explicitly justifiable information in the form of values of that property''. In this work, we refer to the above mentioned empirical property as DVL. The contribution of this review is to bring into light the current stances, limitations, and improvements towards the explicit characterization of DVL as an ``empirical property'' (i.e., existing), and as ``being measurable''.


This study aims at presenting a systematic literature review, with the purpose of making more organic the plethora of material available, of outlining the concerns about DVL, and of answering the following overarching research question: \textit{Which kind of research on DVL characterization and measurement was carried out in the last 10 years?} We answer this question by means of three structured questions, driven by the PRISMA\footnote{PRISMA (Preferred Reporting Items for Systematic Reviews and Meta-Analyses): \url{http://www.prisma-statement.org/}} methodology for systematic literature reviews~\cite{page2021prisma}.

In summary, we are interested in the identification of all of the dimensions that help define, describe, explore, understand, design, model, and assess DVL, namely: i) the terminology and definitions of DVL that were proposed in the literature and what data viz are included, ii) when (year) and where (venue) those researches took place; iii) how the literacy of individuals/data viz difficulty were characterized; iv) whether any model or construct exist; V) whether specific assessment tests were designed, and how they were administered and validated; and, vi) whether other methods than quantitative assessment were employed. To this aim, a query on specific search engines was carried out, resulting in the identification of 675 records. Manual search and snowballing resulted in 43 additional records. After the screening phase, 675 records were excluded, and 43 reports were analyzed in detail. A set of open issues and implications for future research in the DVL topic have been identified, which, together with the analysis of the literature, constitute the main contribution of this paper.

The paper is structured as follows: Section~\ref{sec:related} presents an overview of the related work made in the past about the topic and its satellite aspects; Section~\ref{sec:method} describes the methodology applied for carrying out the systematic review; Section~\ref{sec:results} presents the results of the study and gives answers to the Research Questions in detail; Section~\ref{sec:discussion} discusses the study, in terms of findings, limitations, and research directions; Section~\ref{sec:conc} provides final considerations.


\section{Related Works}
\label{sec:related}

This section describes the literature reviews published in the last 10 years. These reports were mainly related to the antecedents of DVL, its evolution, its objects of study, and its assessment.


\subsection{History and Fortune of the Term Visual Literacy}
One of the first reviews of this last decade was that by~\citet{blummer2015some}, who addressed the broader topic of ``Visual Literacy'' from the point of view of the need of adult individuals to develop visual communications ability, i.e., ``translate from visual language to verbal languages and vice versa'' (ibidem, p.2). In her review, she noted the fact that the field ``lacks a definitive definition of the concept'' (ibidem). She analyzed around 100 reports ranging fifteen years (from 1999 to 2014) of research on Visual Literacy. The review reported some main definitions, perspectives, and studies on Visual Literacy, and then addressed the importance of cultivating Visual Literacy skills in high school and universities. From this main perspective, Blummer identified five educational strategies as outcome of her review, to be pursued through a specialized training activity, and made of several ingredients, including: the need of multidisciplinary and interdisciplinary approaches to Visual Literacy education, of a massive exploitation of multimedia and multi-modal tools, of heterogeneous sources of insights, as well as custom and active participation requests. The use of workshops, tutorials, and projects for students, and of tailored tasks and tests, were crucial facilities for the students' active integration and interventions. The lack of a unified instructional approach was deemed a critical gap at the time of the review. She analyzed reports without classifying them into basic research or applied research, but according to the presence or absence of classes of intervention, such as the presence of custom learning strategies, students materials, courses programs, specific assignments, and the like, for the purpose of identifying the above educational strategies.~\citet{thompson2020uniting} analyzed the literature published from 2011 to 2019 in order to find by whom and to which extent the Association of College \& Research Libraries (ACRL) framework of Visual Literacy Competency Standards was adopted against the proliferation of the many definitions of Visual Literacy. The authors explored the evolution of this framework into a second edition of the ACRL Visual Literacy Competency Standard, published in 2018, and harmonized and integrated into the framework. In this scoping review, there was also a notice related to the open challenges of DVL, which they claimed contains ``widespread disagreements on definitions, measurements, scopes, and terms.'' (ibidem, p.2). 
One of the problems that emerged in this analysis along this time span was that the concept of Visual Literacy itself shifted its focus, following the evolution of the technology of visual means and media. In their analysis of 196 papers, they found that most of the research were conducted in the field of Library Science, followed by Education, Visual Arts, and then STEM and Social Sciences. The terms used in the majority of reports was ``Visual Literacy'', ``Media Literacy'', ``Digital Literacy'' and ``Data Literacy''. Only one report was tagged as ``Data Visualization Literacy''. Visual and performing arts, as well as photography and fine arts, were the main subjects of the identified in the review. They were followed by traditional education topics such as history, STEM subjects, and graphic novels. Although the majority of reports were classified as empirical studies, they noticed that they only mentioned the definition of the ACRL standard, without using the standard approaches proposed in it in the development of their proposals. They also argued that a problem with this standard could have been its western cultural footprint, with no easy possibility of integration into other cultures / nationalities. Another major problem was the adequacy of the standard to the evolution of the discipline and of the enabling technologies.

\subsection{The Usability and Accessibility of Data Viz} 
Although not centered around conceptual stances of DVL, the tutorial review by~\citet{cui2021synergy} touched one of its crucial aspects: data viz as visual representations of statistical concepts. Visual biases of humans were observed as originating from data viz difficulties in correctly vehiculate statistical concepts, derived from the interaction of perceptual objects such as the color, shape, orientation, size, the numerosity of the sample visualized, and the statistical concept to visualize. In their systematization, the authors analyzed around 170 papers\footnote{This number was derived by counting the references, as no summary statistics was provided in the main content of the review about any criteria of searching and inclusion/exclusion}, identified factors that may cause statistical misconceptions in data viz, and suggested what is wrong with data viz. The authors reported the main experiments made in the visual perception fields on data viz (from Cleveland~\cite{cleveland1993model} on). For example, they investigated how the perception of the mean and variance concepts were improved when a wider space was provided for data, instead of crowding data into a tiny space. A sequential presentation of data, where one of the two statistics was kept constant and the other varied progressively was also deemed easier to manage by subjects. Also the kind of granularity of the data helped identify the mean more accurately, e.g., in scatterplots depicting each sample point individually, rather than in bar graphs where subjects tended to underestimate it. Furthermore, the visualization and grasping of the concept of uncertainty were better driven by discrete visual marks, rather than by continuous marks, while confidence intervals or error bars were often misunderstood, with a slight improvement when rendered with animated graphs (the so called HOPs\footnote{Hypothetical Outcome Plots}). Evidences were brought in favor of using different visual features for different tasks: e.g., summarizing data vs. comparing them, organize them into clusters vs. detecting trends, and the like. In conclusion, the authors pointed to proof visual capacity limitations, and tools to spot them. Another review focused on the difficulties of data viz is the one from ~\citet{kim2021accessible}, centered around aspects of accessibility in data viz design, in order to outline how important it was to consider the possible distortions of the visual message, when encoded in the wrong media. In particular, the authors explored the literature of the last 20 years in search of how accessibility was designed, developed and deployed through data viz. An open coding thematic analysis of reports resulted in a final codification along seven principles of design, such as: the background experience of users, the granularity of the data, the type of data viz, the interaction modality, the media used to display the data, the technology and whether it is assistive, and the users' profiling, included levels of disability. This qualitative review explored 56 reports out of an initial selection of 413, of which 38 were classified as artifact based (focused on a new application or technology), 11 as empirical studies, and the few remaining as methodological and reviews. Each of the applications reported were classified and analyzed in relation to each of the seven principles above identified. For example, for each artifact based report, the interaction modalities, the assistive functionality, as well as the assessed impairment of users were reported. The scoping literature review by~\citet{bhat2023infographics} highlighted the importance of using a specific kind of data viz in educational settings. The term infographics referred to visual formats that contain data viz inside, such as bar charts, line charts, and the like. The study related infographics to many dimensions of usability, accessibility and communicability that should also be generalized to data viz. In their review, the authors followed the PRISMA methodology, and analyzed 25 reports, from 2013 to the present days, addressing many related topics of infographics in education. The review developed along some dimensions of learning development that infographics could support. Dimensions include: memory retention, creativity improvement, communication enhancement, and critical thinking exercise. Regarding technicalities of communication the main dimensions were identified in the function of infographics to support storytelling, real time and interactive communication, together with the availability of design guidelines, online templates, and resources for their rapid development. Finally, the authors touched more external topics such as the economic impacts and the limitations devised in all of the reports analyzed relative to the quality, availability, and suitability of infographics in many educational settings, at each level of the learning path of individuals. 

The compendium by~\citet{ansari2022enhancing}, analyzed the state of the art of the applications, framework and evaluations of open data viz technologies. As those data were publicly available, and many important decision were based on them, a glance at their government, especially at how they were represented, summarized  and made available to the citizens is of paramount importance also in the ambit of DVL. In particular, this review took into consideration 39 reports, related to specific data viz applications, and the lessons learnt and challenges for each example reported, the purpose of the data viz and the kind of study, whether an evaluation was carried out for the application at hand, and the like. This first analysis regarded 18 reports. For the frameworks and architectures related to open data government, the same kind of data were reported. This second analysis regarded 8 reports. The description of the stakeholder evaluation and assessment of the systems, when present, were also analyzed. This third analysis regarded 4 reports. A fourth analysis took into consideration the evaluation part of the system / data viz, with details about the number of participants, the kind of evaluation and the findings of each usability study. This analysis regarded 9 reports. Some findings related to the need of engaging more the audience with animated and or decorated data viz, or the need to profile users and provide customized views of the data, and to always provide a user study, were among the main considerations of the study. 

\begin{figure*}[t]
\centering
\includegraphics[width=.7\textwidth]{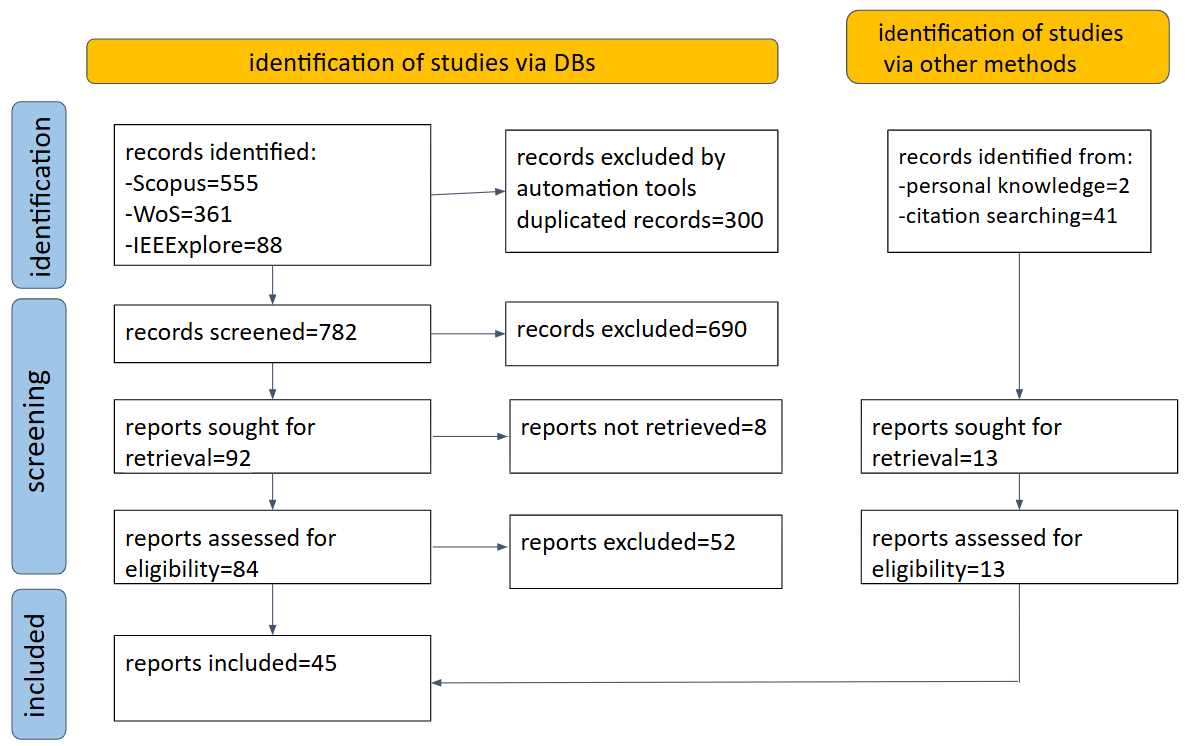}
\caption{PRISMA Flow Diagram followed in this review.}
\label{fig:flow}
\end{figure*}

\subsection{The DVL Assessment} 
The overview by~\citet{ruf2023literature} explored biometric measures to discriminate between the behavior of experts and non-experts users while interacting with data viz. This aspect was related to their proposal of measuring DVL through eye-tracking metrics, as a proxy to measure people's ability to understand data viz. In their review, the authors analyzed 32 reports over the last 10 years, identifying several aspects of the behavioral patterns of people. Diagrams, bar charts, line charts, radar charts, and scatterplots were the data viz analyzed, and the ability dimensions considered were working memory, prio-knowledge, performance, level of study, numeracy, and dislexia. Many tasks were executed by users, for problem solving and learning. The behavioral pattern was discriminant between experts and non-experts, especially in the way, duration, and order of exploration of Areas of Interests (AOIs), properly yielded and identified in data viz. However, the ambiguity and multiplicity of expert vs. non expert users definitions emerged, and four main factors were proposed  for discriminating among the two: data viz knowledge, domain knowledge, mathematical prior knowledge, and task knowledge. The two most recent reviews strictly dealt with the definition and assessment of \textit{DVL}. The two reports were subsequently published by~\citet{firat2022interactive}, and~\citet{firat2022vislite}. The former had a major focus on the state of the art of ``Interactive Visualization Literacy'' from the point of view of the approaches devised in reports published in the main venue of the field: the IEEE VIS conference. The latter was a recap of the former, more focused on the evaluation tests. In these reviews, the reports were categorized based on the kind of evaluations, classified by the same authors with the following keywords: in the wild (i.e., in real world settings), controlled user study, classroom-based, crowdsourced evaluation (e.g., exploiting online platforms, such as MTurk\footnote{\url{https://www.mturk.com/.}}), and literature review. A total of 34 reports were classified according to the above categories, and to the number of participants, to whether they were adults or children, to the kind of task investigated (i.e., reading or writing data viz), and all the data viz used (see the exhaustive Table 7 at page 300 of the first review). The research questions, the limitations, and the approaches used in the reviewed reports to investigate the literacy of participants completed the analysis. Although not explicitly stated in the reports analyzed, many studies were classified as DVL assessment. However, reports about the difficulty of users to find information online (browser search tasks not strictly related to Visuals), as well as experiments with gaze analysis of newspaper pages, and the assessment of the verbal IQ, could be considered as borderline with respect to DVL. Another problem with the two reviews was that they lacked an explicit characterization of DVL. However, many useful and detailed information was available in these reviews: one by one, each assessment was described in a bunch of paragraphs for each report. They made an overview ranging from earlier experimental studies with few data viz, towards more comprehensive frameworks~\cite{Borner20191857}, up to one of the latest validated assessment tests, i.e., the VLAT test~\cite{Lee2017551}. The two reviews were updated to 2022, their publication date. 

According to our best knowledge, no reviews other than those with a specific focus on DVL made extensive considerations about DVL as a property and its foundational utility for the domains, applications, frameworks, kind of data, and data viz proposed.~\citet{firat2022interactive} and~\citet{firat2022vislite} focused on DVL, but they were exclusively centered on its evaluation, neither taking into considerations all the aspects of other reviews nor explicitly following any model or methodology for systematizing their review.   


\section{Method}
\label{sec:method}
Our review adopts a systematic approach, as it allows an overview of the scientific endeavour on the topic of DVL and a deeper investigation on its subtopics, bringing to a robust synthesis for answering our research questions and for outlining the main open issues and future challenges of DVL research. We applied the PRISMA 2020 methodology, included its general approach, guidelines, item checklist, and flow diagram~\cite{page2021prisma}. 
Although PRISMA is primarily designed for clinical studies, and our review concerns a topic more related to Computer Science and Data Visualization, we were able to exploit many items of the methodology as they are. The items used for our work derived from the PRISMA collection are reported in Appendix I. 
Throughout the review, and especially whenever reporting the results and discussing them, we strictly adopt the terminology of PRISMA, and call ``Records'' all the entries with their metadata retrieved from online repositories and selected for inclusion and exclusion, while ``Reports'' are all the papers scrutinized in our review. A list of definitions of PRISMA terms can be found in Appendix II.

\subsection{Research Questions}
\label{sec:rq}
As introduced in Section~\ref{sec:intro}, this systematic review is focused on answering the concerns about DVL characterization and measurement: \textit{Which kind of research on DVL characterization and measurement was carried out in the last 10 years?} To guide our study, we defined a set of specific directions of research:

\begin{enumerate}
\item [RQ1] What are the terms used to identify DVL, its current definitions, and the data viz included?
\item [RQ2] What are the models conceptualizing DVL?
\item [RQ3] What are the empirical studies carried out to characterize DVL?
\item [RQ4] What are the validation and assessment tests for measuring DVL?
\end{enumerate}

\subsection{Search Process}
\label{sec:query}

The Identification phase of the PRISMA method drove us to set criteria for papers search. They were related to the year of publication, namely 2014--2024, to the type of publication (only Journal Articles\footnote{See a detailed justified rationale of this choice in Section~\ref{sec:limitation}}), and to the search engines exploited, namely Scopus\footnote{\url{https://www.scopus.com/search/form.uri?display=advanced}}, Wos\footnote{\url{https://www.webofscience.com/wos/alldb/basic-search}} and IEEExplore\footnote{\url{https://ieeexplore.ieee.org/Xplorehelp/searching-ieee-xplore/advanced-search}}. A template of our search query (in the Scopus syntax) is the following\footnote{the date of last access to the search engines for the query runs was 15th February 2024. We ran the query once again in May 2024 to be sure there were no newly issued articles.}:

\begin{framed}
\footnotesize{\noindent ABS("visualization literacy" OR "visual literacy" OR "visual information literacy" OR "information visualization literacy" OR "data visualization literacy" OR "chart literacy")\\
\noindent OR\\
\noindent TITLE("visualization literacy" OR "visual literacy" OR "visual information literacy" OR "information visualization literacy" OR "data visualization literacy" OR "chart literacy")\\
\noindent OR\\
\noindent TITLE(("visualization literacy" OR "visual literacy" OR "visual information literacy" OR "information visualization literacy" OR "data visualization literacy" OR "chart literacy" OR "information visualization" OR "data visualization" OR "chart") AND "review")\\

\noindent AND NOT TITLE("chart review")

\noindent\begin{equation*}\text{AND DOCTYPE(ar) AND SRCTYPE(j) AND PUBYEAR}  > 2013\end{equation*}}
\end{framed}

We realized that the term ``Chart Review'' was a keyword of the medical domain regarding specific medical studies, which were not relevant for our research, hence we have explicitly asked not to include this term in the papers title.

As a second step, we have merged the results of the above three search queries by discarding duplicated records through the DOI field.

We have added manual sources to the records retrieved by the query, where not present, which were part of the personal knowledge and expertise of the authors, and of snowballing procedures of either queried or manually retrieved works.

The total number of records retrieved after the above procedures was of 675 from queries and 43 from manual search plus snowballing.

\begin{table}[t]
\caption{Records inclusion and exclusion criteria for this study}
\label{tab:incexc}
\centering
\begin{tabular}{|l|}
\hline
\textbf{Inclusion criteria}\\
IC1: Publication date between 2014 and 2024\\
IC2: English language\\
IC3: Focus on Data Visualization Literacy, including elements \\such as: definitions, conceptual modelling, experiments design, assessment,\\
with an emphasis on data viz classification, \\interpretation, and the STEM domains\\
\textbf{Exclusion criteria}\\
EC1: Not meaningful title and abstract\\
EC2: Little or no focus on Data Visualization Literacy\\
EC3: Not in English\\
EC4: Data viz are used but not described nor evaluated\\
EC5: Participants are children of the primary school\\
EC6: Studies focused on data viz tools but not on data viz \textit{per se}\\
EC7: Domain specific data viz (e.g., 3d bio-molecular graphs) \\that are not statistics-based common data viz\\
\hline
\end{tabular}
\end{table}

\subsection{Papers Selection}
\label{sec:selection}

The data retrieved from the query and the merge operation, as well as the manual search were exported to a shared Google Sheet for manual inspection. The records included: the searched DBs, the journal title, publisher, names of the authors, paper title, paper abstract, paper keywords, publication year, DOI, number of citations, link to the source, and a Notes field for free shared comments among the authors.

All four authors checked for inclusion or exclusion of records, according to the inclusion and exclusion criteria depicted in Table~\ref{tab:incexc}. The procedure of record selection was completed in three steps. In the first step, two rounds of ``calibration'' took place. In the first round, each author classified individually the first five records of the sheet. This first phase brought to a disagreement of three records out of five that was solved after discussion. The subsequent five records in the second round were also managed separately, and an agreement discussion was necessary on only one disagreement. After this session, the second step regarded the assignment to each author of a quarter of all the remaining records for individual classification into three categories: ``keep'', ``discard'', and ``discuss'' (this last category was necessary whenever the author was not sure about the keeping or discarding decision of the record, and a discussion with the other authors was needed). A group session to definitely keep or discard the ``discuss'' records was then carried out. Around one third of the records classified by each author were initially put in the ``discuss'' category. All of them were kept or discarded  after discussion among all of the authors. The most part of these records were discarded in the end. As a double-check, each author revised once again one quarter of the records discarded, among the ones not discarded personally, and could promote them back to ``discuss'' for another discussion round and ultimate decision about the paper. Very few records were discussed once more in this phase, and only half of them were promoted to ``keep''. The third step of the selection regarded the scrutiny of all the records labelled with ``keep'' for data extraction.

\subsection{Data Extraction}
\label{sec:extraction}

For the scrutiny of the records kept, two calibration rounds were first of all conducted on the first five records and again on the subsequent five records of the common spreadsheet (each author carried out the data extraction individually and then a common session served for the final discussion). The data extraction items are listed in Table~\ref{tab:ext}. These extraction items were put as further columns in the spreadsheet and each author had to fill them for each of the records assigned. A copy of the sheet was necessary to keep this process separate for each author. A sampling session of ten records was discussed together on the work done individually, to be sure that each data extracted by the single author was consistent among the authors.  

After this calibration phase, each author was assigned a quarter of the records selected for final scrutiny, the reports were downloaded in pdf and put in a directory named with the name of each author. 
Then the final data for all included records were copied back again in the shared spreadsheet. 

\section{Results}
\label{sec:results}
Figure~\ref{fig:flow} shows the complete flow diagram of the records and reports Selection. After the automatic deletion of duplicate records and a first screening phase by title and abstract reading, we obtained a number of 111 records sought for retrieval, out of which 98 were retrieved from the search engine query, and 13 were identified among those manually retrieved thanks to the expertise of some of the authors and of snowballing procedures. Eight records were excluded from this selection due to exclusion criteria EC1 and EC3 (i.e., their title and/or abstract was not meaningful of their content, and they were not in English language). All the 82 remaining records from search queries and all the 13 manually selected records were downloaded and assessed for eligibility. After reading them, 52 of them were excluded because they were not pertinent to the study (see exclusion criteria EC2, EC4, EC5, EC6, AND EC7 in Table~\ref{tab:incexc}). As done for the initial task, also for the session of records eligibility and scrutiny, the downloaded reports were equally divided by the four authors, further and separately analyzed for discussion, so that a final agreement could be reached to exclude or include reports, until 43 of them were definitely included for scrutiny. 


\begin{table}[t]
\caption{Data extracted from the scrutiny of reports}
\label{tab:ext}
\centering
\begin{tabular}{|l|}
\hline
DE1: Year\\
DE2: Journal, Journal IF and Publisher\\
DE3: \begin{tabular}[t]{@{}l@{}}Type of paper: Basic research (creation/extension\\ of models of Data Visualization Literacy, creation/extension \\of models of data visualization classification \\or difficulty rankings, creation/extension of \\assessment tests); Literature Review (review)\footnote{Review of the kind explicitly defined and described at \url{here: https://guides.mclibrary.duke.edu/sysreview/types}}; \\Applied research (empirical studies comparing \\existing data viz, user studies)\end{tabular}\\
DE4: Definitions of Data Visualization Literacy (and related terms) \\
DE5: Domain (healthcare, avionics, education, and the line)\\
DE6: Participants (students, professionals, lay people) \\
DE7: Data viz usability (yes/no)\\
DE8: Persons ability (yes/no)\\
DE9: Main purpose of the contribution \\(model, questionnaire, and the like)\\
DE10: Data viz used / evaluated / designed\\ 
DE11: \begin{tabular}[t]{@{}l@{}}Type of statistical assessment \\(name of the model /test, items type and \\number, number of respondents)\end{tabular}\\
\hline
\end{tabular}
\end{table}

An overview of the reports, with authors, year of publication, journal, type of report, domain of interest, sample population involved, main contribution, and availability of an assessment questionnaire is reported in Table~\ref{tab:overview}. Among the 43 reports analyzed, quite a half (19) proposed a new questionnaire (or a shorter version of an existing one) or exploited an existing assessment test. The majority of the reports were classified as basic research, as they proposed a conceptualization of DVL, mainly based on existing theories from the learning domain, and having as outcome a list of DVL skills. A third of the reports with an assessment test used MTurk, and four other reports used the Prolific portal. This online-based experiments were exploited in half of the reports with an assessment test. The majority of reports with an assessment test declared to use students as participants of their experiments. Many reports (16) focused on the education domain, and 21 referred to the population in general. Three reports came from the specific domain of health, witnessing a lively community of scholars interested in the topic of DVL for patients and physicians. Conceptual only reports for the DVL were 18, while 9 proposed only a questionnaire, and  7 reports presented both. 

\begin{figure}[t]
\centering
\includegraphics[width=.485\textwidth]{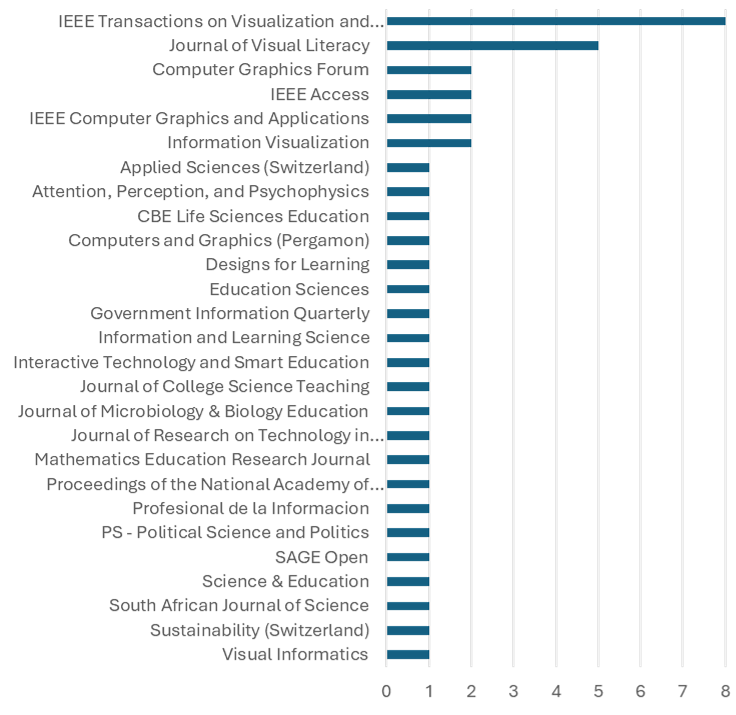}
\caption{The distribution of the 43 reports considered in this systematic review by journal title.}
\label{fig:distJ}
\end{figure}

\begin{figure}[t]
\centering
\includegraphics[width=.485\textwidth]{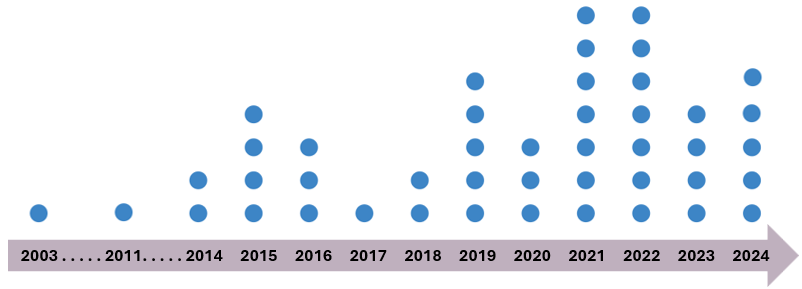}
\caption{The distribution of the 43 reports considered in this systematic review by publication date.}
\label{fig:time}
\end{figure}

Figure~\ref{fig:distJ} depicts the distribution of the analyzed reports by Journal title, in descending order, whereas the time span of publication of the reports analyzed is reported in Figure~\ref{fig:time}.  


\begin{table*}[t]
\caption{Overview of the 43 reports analyzed in this systematic review.}
\label{tab:overview}
\centering
\begin{threeparttable}
\begin{tabular}{|l|c|l|l|l|l|l|c|}
\hline
Authors & Year & Journal & Type$^{a}$ & Domain$^{b}$ & Sample$^{c}$ & Main$^{d}$ & Quest \\ \hline
Bhat, Alyaya & 2024 & IEEE Access & review & education & - & - & \\
Cui et al. & 2024 & IEEE Trans on Visualiz \& Comput Graphics & basic & general & general \mycircle{black} & quest & \checkmark\\
Davis et al. & 2024 & IEEE Trans on Visualiz \& Comput Graphics & basic & general & general \mytriangle{black} & model & \checkmark \\
Kalaf-Hughes & 2023 & Political Science and Politics & basic & university & student & model & \\
Pandey, Ottley & 2023 & Computer Graphics Forum & basic & general & general \mycircle{black} & quest & \checkmark\\
Reddy et at. & 2023 & Interactive Technology and Smart Education & basic & high-school & student & mod \& q & \checkmark\\
Ruf et al. & 2023 & Education Sciences & review & general & - & - & \\
Ansari et al. & 2022 & Government Information Quarterly & review & general & - & - & \\
Binali et al. & 2022 & Science \& Education & applied & education & student & model & \checkmark\\
Camba et al. & 2022 & IEEE Computer Graphics and Applications & applied & university & student & model & \checkmark \\
Firat et al. & 2022 & Visual Informatics & basic & general & general \mytriangle{black} & quest & \checkmark \\
Firat et al. & 2022 & IEEE Computer Graphics and Applications & review & general & - & - & \\
Firat et al. & 2022 & Information Visualization & review & general & - & - & \\
Mart{\'\i}n Erro et al. & 2022 & Journal of Visual Literacy & basic & engineering & - & model & \\
Cui, Liu & 2021 & Attention, Perception, and Psychophysics & review & general & - & - & \\
Kim et al. & 2021 & Computer Graphics Forum & review & general & - & - & \\
Locoro et al. & 2021 & IEEE Access & basic & general & general \mycircle{black} & mod \& q& \checkmark\\
Lundgard, Satyan. & 2021 & IEEE Trans on Visualiz \& Comput Graphics & basic & general & general \mycircle{black} & model & \\
M.B. Rodrigues et al. & 2021 & Computers and Graphics & applied & unversity & student & model & \\
Peppler et al. & 2021 & Information and Learning Science & applied & museum & general & model & \\
Yang et al. & 2021 & IEEE Trans on Visualiz \& Comput Graphics & applied & general & general \mycircle{black} & mod \& q & \checkmark\\
Krejci et al. & 2020 & Sustainability & basic & university & student & mod \& q& \checkmark\\
Marzal & 2020 & Profesional de la Informacion & basic & general & general & model & \\
Thompson, Beene & 2020 & J. of Visual Literacy & review & general & - & - & \\
Bod{\'e}n, Stenliden & 2019 & Designs for Learning & applied & high-school & student & model & \\
B{\"o}rner et al. & 2019 & National Academy of Sciences of the US & basic & general & - & model & \\
Lee et al. & 2019 & Applied Sciences & basic & general & general \mytriangle{black} & mod \& q& \checkmark\\
Okan et al. & 2019 & Medical Decision Making & basic & health & academic & quest & \checkmark\\
Stenliden et al. & 2019 & J. of Research on Technology in Education & applied & secondary school & student & model& \\
Arneson, Offerdahl & 2018 & CBE Life Sciences Education & basic & biology univ & student & model & \\
K{\c e}dra & 2018 & J. of Visual Literacy & basic & general & - & model & \\
Lee et al. & 2017 & IEEE Trans on Visualiz \& Comput Graphics & basic & general & general \mytriangle{black} & mod \& q& \checkmark\\
B{\"o}rner et al. & 2016 & Information Visualization & applied & museum & general & model & \\
Garcia-Retamero et al. & 2016 & Medical Decision Making & basic & health & general \mytriangle{black} & quest & \checkmark\\
Mnguni et al. & 2016 & South African J. of Science & basic & biochemistry univ & student & model & \\
Blummer & 2015 & J. of Visual Literacy & review & education & - & - & \\
Bresciani, Eppler & 2015 & SAGE Open & basic & data viz design & - & model & \\
Maltese et al. & 2015 & J. of College Science Teaching & basic & university & student & quest & \checkmark\\
Vermeersch, Vandenb. & 2015 & J. of Visual Literacy & basic & high-school & student & model & \\
Boy et al. & 2014 & IEEE Trans on Visualiz \& Comput Graphics & basic & general & general \mytriangle{black} & quest & \checkmark\\
Rybarczyk et al. & 2014 & J. of Microbiology \& Biology Education & basic & biochemistry univ & student & quest & \checkmark\\
Galesic, Garcia-Retamero & 2011 & Medical Decision Making & basic & health & general & mod \& q & \checkmark \\
Aoyama, Stephens  & 2003 & Mathematics Education Research J. & applied &general & general & quest & \checkmark \\
\hline
\end{tabular}
\begin{tablenotes}\scriptsize

\item[a] \textit{Type} refers to our classification: basic research (basic), applied research (applied), and literature review (review).
\item[b] \textit{Domain} is the domain of the report, or general if no domain is given. When further detailed, the educational level has been reported, otherwise the term education is used instead. 
\item[c] \textit{Sample} contains the kind of sample used in the report, if any. A circle stands for \textit{Prolific}, the triangle stands for \textit{MTurk} of samples.
\item[d] \textit{Main} shows the main contribution of reports: \textit{model}, for construct/set of skills/model of DVL; \textit{quest}, for questionnaire; \textit{mod\&q} for both.  
\end{tablenotes}
\end{threeparttable}

\end{table*}

\subsection{RQ1. Terms and Definitions for DVL}

As evidence of the heterogeneity introduced in Section~\ref{sec:intro}, and impossibility to exclude any term or definition from our horizon of inquiry, we report a complete overview of the terms exploited to determine DVL and their definition, as an outcome of our work. Table~\ref{tab:definitions} recollects them and updates the older study similar to ours in this respect~\cite{locoro2021visual}.

\begin{table*}[ht]
\footnotesize
\caption{An overview of the different terms and definitions for \textit{DVL}.} 
\label{tab:definitions}
\begin{tabular}{ll} \\\toprule
Expression &
  Definition \\\hline
\begin{tabular}[t]{@{}l@{}}Visual literacy \\
~\citet{doi:10.1080/15391523.2018.1564636}\end{tabular} &
\begin{tabular}[t]{@{}l@{}} the ability to retrieve information in a visual analysis process paired with the ability to express the \\ resulting knowledge through visual messages. \end{tabular}\\\hline
\begin{tabular}[t]{@{}l@{}}Visual literacy \\
~\citet{kalaf2023promoting}\end{tabular} &
  \begin{tabular}[t]{@{}l@{}}The ability to successfully and ethically engage with visual media and under-
stand \\ how it is produced and valued\end{tabular} \\\hline
\begin{tabular}[t]{@{}l@{}}Visual literacy\\ ~\citet{reddy2023visual}\end{tabular} &
  \begin{tabular}[t]{@{}l@{}}the ability of a person to work with images.\end{tabular} \\\hline
\begin{tabular}[t]{@{}l@{}}Visual literacy\\ ~\citet{boden2019emerging}\end{tabular} &
  \begin{tabular}[t]{@{}l@{}}as the ability to understand the composition and meaning of a visual property through interpretation and \\analysis, are enacted in an educational setting.\end{tabular}  \\\hline
\begin{tabular}[t]{@{}l@{}}Visual literacy \\ ~\citet{vermeersch2015kids}\end{tabular} &
  \begin{tabular}[t]{@{}l@{}}a process based on a number of (nonverbal) cognitive skills. Visual literacy is more than seeing images \\and making images (such as photos, paintings, drawings, etc.) as natural or neutral activities; \\it is a thinking process, one that implies specific cognitive actions like interpretation and reflection, \\understanding and comprehension, awareness, etc.\end{tabular}  \\\hline
\begin{tabular}[t]{@{}l@{}}Visual Literacy Competency Standards \\ (ACRL~\cite{thompson2020uniting})\end{tabular} &
  \begin{tabular}[t]{@{}l@{}}A set of abilities that enables an individual to effectively find, interpret, evaluate, use, \\and create images and visual media. Visual literacy skills equip a learner to understand and \\ analyze the contextual, cultural, ethical, aesthetic, intellectual, and technical components \\involved in the production and use of visual materials. A visually literate individual \\is  both a critical consumer of visual media and a competent contributor to a body \\ of shared knowledge and  culture.\end{tabular} \\\hline
\begin{tabular}[t]{@{}l@{}}Visualization Literacy (VL)\\ ~\citet{lee2019correlation}\end{tabular} &
  \begin{tabular}[t]{@{}l@{}}the ability and skill to read and interpret visually represented data in \\
and to extract information from data visualizations\end{tabular} \\\hline
\begin{tabular}[t]{@{}l@{}}Visualization literacy\\ ~\citet{firat2022interactive}\end{tabular} &
  \begin{tabular}[t]{@{}l@{}}an essential skill required for comprehension and interpretation of complex imagery
\\ conveyed by interactive visual designs.\end{tabular}  \\\hline
\begin{tabular}[t]{@{}l@{}}Visualization literacy\\ ~\citet{cui2023adaptive}\end{tabular} &
  \begin{tabular}[t]{@{}l@{}}An individual’s ability to understand and interpret visualizations\end{tabular} \\\hline
\begin{tabular}[t]{@{}l@{}}Visualization literacy\\ ~\citet{borner2016investigating}\end{tabular} &
  \begin{tabular}[t]{@{}l@{}}the ability to make meaning from and interpret patterns, trends, and correlations \\in visual representations of data.\end{tabular}  \\\hline
\begin{tabular}[t]{@{}l@{}}Visualization literacy\\ ~\citet{6875906}\end{tabular} &
  \begin{tabular}[t]{@{}l@{}}the ability to use well-established data visualizations (e. g., line graphs) \\to handle information in an effective, efficient, and confident manner.\end{tabular}  \\\hline
\begin{tabular}[t]{@{}l@{}}Data literacy (DL)\\ ~\citet{marzal2020taxonomic}\end{tabular} &
  \begin{tabular}[t]{@{}l@{}}understanding of information represented by numbers in the broadest sense, together
with \\the information used by algorithms and that can be presented visually\end{tabular}  \\\hline
\begin{tabular}[t]{@{}l@{}}Data Visualization literacy (DVL)\\
~\citet{rodrigues2021questions}\end{tabular} &
  \begin{tabular}[t]{@{}l@{}}the ability to use well-established data visualizations (e.g., line graphs) to handle \\information in an effective, efficient, and confident manner.\end{tabular}  \\\hline
  \begin{tabular}[t]{@{}l@{}}Data Visualization literacy\\ ~\citet{peppler2021cultivating}\end{tabular} &
  \begin{tabular}[t]{@{}l@{}}the ability to read and construct visual representations to make meaning of data and \\to support the understanding of datasets through data visualization types (e.g. scatter graph, geo map), \\data variables (i.e. qualitative, quantitative) and graphic variable types (e.g. shape, size, color).\end{tabular}  \\\hline
\begin{tabular}[t]{@{}l@{}}Parallel Coordinate Plot literacy \\
~\citet{firat2022p}\end{tabular} &
  \begin{tabular}[t]{@{}l@{}}the ability to correctly read, interpret, and construct PCPs.\end{tabular}  \\\hline
\begin{tabular}[t]{@{}l@{}}Visual Information literacy (VIL)\\~\citet{locoro2021visual}\end{tabular} &
  \begin{tabular}[t]{@{}l@{}}the ability to properly process information related to data viz, i.e., \\encoding information into data viz and decoding information from data viz\end{tabular} \\\hline
\begin{tabular}[t]{@{}l@{}}Graph Literacy (GL)\\
~\citet{Galesicetal2011}\end{tabular} &
  \begin{tabular}[t]{@{}l@{}}the ability to understand graphically presented information\end{tabular} \\\bottomrule
\end{tabular}
\end{table*}

\subsection{RQ2. Basic Research Reports}
\label{sec:basic}
Basic research reports are those proposing a characterization of DVL and its dimensions (e.g., an explicit model, the skills as dimensions of the model, and the like), and the design of an assessment, where present. This second element is treated in Section~\ref{sec:assess}, where RQ4 is addressed. 

An overview of models, dimensions and skills proposed or exploited in the reports analyzed is reported in Tables~\ref{tab:skill1} and~\ref{tab:skill2}. The following analysis of the reports has been grouped into three main strands: defining and promoting DVL for education and learning; proposing constructs and frameworks for characterizing and assessing DVL; and exploiting models of cognition and perception in order to study DVL in adults and impaired people.

\subsubsection{Defining, Promoting and Supporting DVL skills}

Many reports described the concept of DVL by identifying the essential skills that individuals must attain to be considered as visually literate. Starting from the farthest domain with respect to data visualizations, that of visual arts, along the multiliteracies of social, information and communication media, we touched upon the conceptualizations devised for the STEM domains towards DVL as a general property of common people interacting with data viz. 

\citet{vermeersch2015kids} expanded upon Van Heusden's theory, which outlined the stages of semiosis in visual arts. The first skill characterizing their model was \emph{visual perception}, which involves cultivating a ``good eye'' in experiencing the visual, including being curious, sensitive, focused, selective, and deriving pleasure from it. \emph{Visual imagination and creation} and \emph{conceptualization} involve using perceived images to generate something new, based on the capacity to understand image structure like a symbolic language made of visual elements (points, lines, angles, open and closed figures), and transforming concrete perceptions into abstract representations through abstraction, thereby converting them into conceptual signs or symbols. 
\emph{Analysis of images} implies identifying structures or patterns within data viz. Unlike conceptualization, analysis is a systematic and structured approach that does not rely on social conventions. It entails an orderly exploration of visual stimuli to uncover underlying structures and meanings to support visual reasoning and thinking.

\citet{Kedra201867} defined skills related to the context of visually mediated communication, digital technologies, and new media.
The report contributed to the topic by examining various definitions of DVL and organizing them into three skills: \textit{visual reading} skills (interpreting, meaning making), \textit{visual writing} skills (using or creating images), and \textit{visual thinking and learning} abilities. 
Other DVL skills involved \textit{visual thinking}, \textit{visual learning}, and \textit{image use}, which should develop along the process of visual education. Competency in visual thinking was cultivated through visual education, where learners engaging with images systematically enhanced their visual perception.  The report by~\citet{doi:10.1128/jmbe.v15i2.703} focused on developing DVL skills in scientific literacy for undergraduate science education. Data analysis skills were identified and applied to the world of learning in the biological sciences, but they were deemed as generalizable to the scientific reasoning and interpreting of scientific data. The authors asked some DVL experts to identify and compile a list of essential skills necessary for students to interpret data in visual formats.
The experts identified some \emph{basic level} skills, fundamental for describing the surface-level characteristics of a visual, and an \emph{advanced level} of skills, needed for the analysis, inference, and interpretation of contextual information with reasoning abilities. The multiliteracy constellation proposed by~\citet{marzal2020taxonomic} also highlighted the connection and interdependence of various literacies related to DVL: digital literacy, information literacy, and media literacy, among others. Integrating those literacies into a metamodel was the research proposal of these reports, investigating the DVL property in the broader spectrum of educational perspectives. 

\begin{table*}[h!]
\centering
\caption{DVL models and their dimensions, with detailed DVL skills of the analyzed reports for strand 1: defining, proposing and supporting DVL skills.}
\label{tab:skill1}
\begin{tabular}{|l|l|l|}
\hline
Report & Model Dimensions / Levels & Skills \\ \hline
~\citet{vermeersch2015kids} & perception &  -\\ 
 & imagination and creation & -\\ 
 & conceptualization &  -\\ 
 & analysis &  -\\ \hline
~\citet{Kedra201867} & visual reading & interpret, analyze, understand, evaluate, perceive \\ 
 &  & know grammar and syntax \\ 
 &  & translate (visual-verbal-visual) \\ 
 & visual writing & communicate visually \\ 
 &  & create and produce images \\ 
 &  & use \\ 
 & other visual literacy skills & think visually \\ 
 &  & learn visually \\ 
 &  & apply images \\ \hline
~\citet{doi:10.1128/jmbe.v15i2.703} & basic-level & identify patterns and trends in data \\ 
 &  & connect source of the data \\ 
 & advanced-level & distinguish between positive and negative controls, propose other controls \\ 
 &  & synthesize conclusions from data\\ 
 &  & determine whether data supports a hypothesis \\ 
 &  & propose follow-up experiments, predict results \\ \hline
~\citet{arneson2018visual} & knowledge & memorize, recognize, recall, retrieve \\
 & comprehension & understand, interpret, infer, exemplify, classify \\ 
 & application & execute, implement, apply \\ 
 & analysis & differentiate, discriminate, organize, integrate, deconstruct, attribute \\ 
 & evaluation & check, coordinate, critique, judge, test \\ 
 & synthesis & generate, hypothesize, plan, design, construct, reorganize, produce \\ \hline
~\citet{kalaf2023promoting} & information literacy & retrieve, organize \\ 
 & visual literacy & perceive, analyze, represent, use multi-modality \\ 
 & critical thinking & think analytically, for problem-solving, for decision-making \\ 
 & communication & communicate effectively, collaborate with peers, speak publicly \\ 
 & adaptability and lifelong learning & adapt, learn lifelong \\ \hline
~\citet{mnguni2016assessment} & internalisation & arrange, order, organise, classify \\ 
 &  & recognize, find, locate, focus, ground \\
 &  & perceive visual stimuli such as luminance, color, and  motion \\ 
 & conceptualisation & analyse, interpret, assess, evaluate, examine, investigate \\ 
 &  & compare, relate \\ 
 &  & critique, imagine, describe, discuss, explain, discriminate, judge \\ 
 &  & manipulate objects using geometrical properties and operations \\
 &  & recall, retrieve \\
 & externalisation & complete \\ 
 &  & illustrate, sketch \\ 
 &  & infer, predict \\ 
 &  & outline \\ 
 &  & propose, develop, formulate, devise, construct, create, produce, invent \\ 
 &  & use \\ 
 &  & define the need for an image \\ \hline
~\citet{martin2022framework} & seeing stage & use and interpret \\ 
 & imagine stage & analyze and solve \\ 
 & drawing stage & create and communicate \\ \hline 
~\citet{reddy2023visual} &  & find images \\ 
 & - & interpret and analyze images \\ 
 & - & evaluate images \\ 
 & - & use images effectively \\ 
 & - & create visual media \\ 
 & - & use images ethically and citing visuals \\ \hline
~\citet{krejci2020visual} & elementary & interpret \\ 
 & intermediate & reason, explain, evaluate \\
 & advanced & synthesize, reflect, judge \\ \hline
\end{tabular}
\end{table*}

\citet{arneson2018visual} and~\citet{mnguni2016assessment} argued how visual representations play a crucial role in science communication, and how promoting scientific visual literacy should be a key goal for undergraduate teaching. Both reports relied on Bloom's Taxonomy\footnote{Available at: \url{https://cft.vanderbilt.edu/guides-sub-pages/blooms-taxonomy/.}}. The former report modified it into a Visualization Blooming Taxonomy Tool (VBT) that teachers could use to classify visual questions based on cognitive levels and create assessments aligned with DVL objectives. For each cognitive level the authors outlined a range of skills: \emph{knowledge} involves memorizing, recognizing, and recalling data viz, like identifying image components and defining symbols; \emph{comprehension} includes interpreting visual representations to extract relevant information and make predictions; \emph{application} requires using procedures to solve problems with visual data, possibly modifying processes; \emph{analysis} involves evaluating data viz by breaking it down into smaller parts and identifying relationships; \emph{synthesis} combines data viz to create new representations, while \emph{evaluation} involves critically assessing visual representations validity and effectiveness. The latter report relied on a simplified version of the Bloom's taxonomy, where the three stages of internalization, conceptualization, and externalization are described and assessed through a test with 24 biochemistry visualizations. The dimension of \emph{internalization} involves absorbing external information through sense organs, while that of \emph{conceptualization} entails constructing cognitive visual models. During conceptualization, existing knowledge is retrieved from long-term memory and modified based on new information. \emph{Externalization} expresses cognitive mental schema through the production of external visual models.~\citet{kalaf2023promoting} discussed the importance of teaching Information Literacy and DVL skills to students in both academic and professional areas. They proposed an experiment where students had to translate a scholarly article into an infographics and participate in a discussion to reinforce their DVL skills. They devised the following DVL skills: \textit{information literacy} to research information, evaluate its quality, and effectively communicate scholarly information to peers by synthesizing and organizing information; \textit{visual literacy}, to interpret data viz, create and design visual content, and understand how visuals can effectively communicate complex information; 
\emph{citical thinking}, to mature an analytical mindset, problem solving and decision making attitudes; \emph{communication}, to develop effective communication, peer collaboration, and public presentation abilities; \emph{adaptability and lifelong learning}, to possess adaptable skills that can be applied across disciplines and in various professional settings, instilling a mindset of continuous learning to thrive in a rapidly changing academic and professional landscape.

Similarly,~\citet{martin2022framework} proposed a ``Competence Framework'' to develop visual skills for thinking, communicating, and learning, tailored for the specific needs in the engineering curricula. 
The project incorporated visual thinking principles based on Robert McKim's framework~\cite{von2019theoretical}, which categorizes visual imagery into what is seen, imagined, and expressed through drawings, in a cyclical process termed ``seeing-imagining-drawing.''. The VLEE-CF framework integrated this cyclical approach into its six competence areas, linking ``use'' and ``interpretation'' with the initial \textit{seeing stage}, ``analysis'' and ``solving'' with the \textit{imagine stage}, and ``creation'' and ``communication'' with the final \textit{drawing} stage. Three levels for each stage were devised: \textit{basic} (assimilation of new information alongside the execution of elementary practices), \textit{intermediate} (application, extension, and reflection of these practices), and \textit{proficient} (knowledge transmission, critique of existing practices, and the development of new materials).

\textit{Critical thinking} is also a relevant skill for DVL in education.~\citet{reddy2023visual} and~\citet{krejci2020visual} outlined those aspects in their list of skills. Students' understanding were encouraged by facilitating group discussion and active participation. A tripartite model was proposed, with \textit{elementary}, \textit{intermediate}, and \textit{advanced} levels used to evaluate and monitor the degree of visual competence of students and to measure progress in improving their skills. The elementary level focuses on the student's ability to read and understand data presented in a visualization, the intermediate level requires students to analyze relationships between the information presented, and the advanced level requires students to draw conclusions or make predictions.

\subsubsection{Constructs for assessing the DVL property}
Very few studies started with the identification of DVL skills as a foundational step towards defining a construct map for DVL. This process served as a precursor to developing competency assessments aimed at evaluating individuals' proficiency in DVL, considering data viz. For example,~\citet{locoro2021visual} propose a definition of DVL and designed a model that characterizes it as a developmental progression of skills.
The report described the systematic approach used to answer questions related to data viz difficulty and graph-supported task complexity. The DVL model is based on Piaget's theory for effective data viz processing. It consists of seven levels based on people's DVL degree. The levels of progression in DVL are: to be able to \textit{map} visual syntax into visual information properties; \textit{integrate} it into a whole meaning; \textit{compute} the quantitative information presented in data viz; \textit{reason} to derive conclusions from the presented information; \textit{infer} new information and knowledge; be able to \textit{explain} and critically judge, evaluate and debunk visual information. Tasks like understanding visual syntax, interpreting data viz, analyzing patterns, and the like, are mirrored into items for assessing the level of attainment of users. 

\citet{Borner20191857} presented a framework for defining, teaching, and assessing skill highlighting the need to include exercises and assessments in existing tests to evaluate how effectively students were prepared to read and construct data viz.
The key components of their  Data VL Framework (DVL-FW) included several key aspects related to the hierarchical typology of core concepts associated with DVL, and a detailed outline of the steps necessary to derive insights from data viz.
The \textit{typology} elements included insights, scales, kinds of analyses (statistical, temporal, topical, relational), kind of data viz, used symbols (geometric, linguistic, pictorial), encoding variables (spatial, retinal), and interaction modes (e.g., zoom, search, filter, details on demand, history).
The authors derived a \textit{step-wise} process model for constructing and interpreting data viz.
The first step was the identification of stakeholders' insights, relevant datasets, data preprocessing and visualization, deployment, and interpretation of results. The iterative nature of the process often leads to refining data, analysis, visualization workflows, and deployment options based on the insights gained. Different deployment methods and interface controls supported various interactions. The final step involved translating visualization results into actionable insights for real-world applications.
The DVL-FW process model guided students through the iterative design of data viz, with assessments evaluating their ability to create, interpret, and evaluate data viz effectively. Assessments focused on interpreting data viz across different tasks and DVL-FW types, aiming to identify and address misinterpretations. 

\begin{table*}[ht]
\centering
\caption{Models and Constructs, with detailed dimensions of the analyzed reports for strand 2 and 3: constructs for assessing the DVL property, perceptual and cognitive models for adults }
\label{tab:skill2}
\begin{tabular}{|l|l|l|}
\hline
Report & Model / Construct Type & Dimensions \\ \hline
~\citet{Borner20191857} & DVL- Framework typology & insight needs, data scales, analyses, visualizations \\ & & graphic symbols, graphic variables, interactions \\
& DVL-Framework process model & stakeholders, acquire, analyze, visualize \\ & & deploy, interpret \\ \hline

~\citet{bresciani2015pitfalls} & visualizations pitfalls &  \textit{encoding (designer-induced)} \\ 
& & - cognitive: ambiguity, breaking conventions, confusion, \\ & & cost to make explicit, cryptic encoding, defocused, \\ & & hiding/obscuring, inconsistency, low accuracy,  \\ & & misleading/distorting, not respecting gestalt principles,  \\ & & over-determinism, over/under-reliability appearance, \\ & & 
over-complexity, over-simplification, 
 redundancy,\\ & &  task-visualization fit, technology/template driven,\\ & & time-consuming to produce, unclear, unevenness \\
 & & - emotional: disturbing, boring, ugly, wrong use of color \\ & & 
 - social: affordance conflict, hierarchy, exercise of power, \\ & & inhibit conversation, rhythm of freezing and unfreezing, \\ & & turn-taking alteration, unequal participation \\ 
& & \textit{decoding (user-induced)} \\ & & 
- cognitive: change blindness, channel thinking, \\ & & depending on perceptual skills, \\ & & 
difficult to understand, focus on low relevance items, \\ & & high requirement on training and resources, \\ & & 
knowledge of visual conventions, misuse, overload, \\ & & reification, wrong salience \\ & & 
- emotional: visual stress, personal likes and dislikes, \\ & & prior knowledge and experience \\ & & 
- social: altered behavior, cultural and cross-cultural differences, \\ & & defocused
from non-verbal interaction, different, \\ & & hiding differences of opinion, recency effect, \\ & &  time-consuming to agree
\\ \hline

~\citet{davis2024risks} & visualization tasks &  estimate and compare values \\ &  & judge correlations \\ &  & identify trends and outliers \\ &  & spot suitability  \\ \hline
~\citet{Galesicetal2011} & graph literacy levels &  1st level: read the data \\ &  & 2nd level: read between the data \\ &  & 3rd level: read beyond the data \\ \hline
~\citet{lee2019correlation} & cognitive characteristics &  numeracy, need for cognition, visualizer-verbalizer capacity \\ \hline

~\citet{locoro2021visual} & construct levels of VIL &  map, integrate, compute, reason, infer, explain \\ \hline

~\citet{lundgard2021accessible} & model levels of DVL &  enumerating visualization construction properties \\ &  & reporting statistical concepts and relations \\ &  & identifying perceptual and cognitive phenomena \\ &  & elucidating domain-specific insights  \\ \hline
\end{tabular}
\end{table*}

\subsubsection{Perceptual and Cognitive Models for Adults}
Many reports discussed the importance of perceptual and cognitive characteristics of individuals and accessibility of data viz in this respect. They focused on evaluating people's performance based on data viz tasks, in order to help contextualize data viz design approaches. 
For example,~\citet{davis2024risks} showed how significant the variability of individuals' performance interacting with data viz, with respect to an ``average user'' was. 
In this regard, some reports focused on analyzing which individual characteristics came into play in reading data viz.~\citet{lee2019correlation} suggested three cognitive characteristics: \textit{numeracy}, \textit{need for cognition}, and \textit{visualizer-verbalizer style}. Numeracy was defined as the ability to understand numerical information; the need for cognition was related to an individual's tendency towards engaging in cognitive activities; the visualizer-verbalizer style referred to preferences for visual or verbal information processing. The tested hypothesis was that different degrees of each characteristic may have impacts on how readers extracted more information accurately from data viz. It was found that numeracy had a positive correlation with DVL. On the other hand, individuals with low numeracy were found to struggle with visualization tasks requiring numerical manipulation, for lack of understanding of visual encoding schemes and a tendency to overlook visually represented numerical data. As also mentioned in~\citet{davis2024risks}, individuals' differences were deemed essential in creating guidelines for data viz designers.~\citet{bresciani2015pitfalls} discussed the various errors and challenges that could arise in the design and interpretation of information and knowledge visualizations. The authors provided a classification of common errors into two categories: ``encoding'' (designer-induced), and ``decoding'' (user-induced). The main pitfalls were then categorized into three types, according to their negative effects: \textit{cognitive}, \textit{emotional}, and \textit{social}. Some examples of included pitfalls are: ``ambiguity'', ``confusion'', ``overload'', ``emotional disturbance'', and ``cultural differences''. 
Another important aspect linked to the difficulty in reading data viz concerned accessibility issues.~\citet{lundgard2021accessible} addressed the evaluation of natural language descriptions accompanying data viz in enhancing design and conveying meaningful information to readers, especially those with disabilities. To this aim, they introduced a four level conceptual semantic model, using grounded theory, to evaluate accessibility of data viz: \textit{level 1} describes basic visual components like chart type and axis labels; \textit{level 2} covers statistical concepts and relations from the dataset; \textit{level 3} focuses on complex trends and patterns, requiring human perception for interpretation; \textit{level 4} provides contextual and domain-specific insights, relying heavily on individual expertise.


Finally,~\citet{Galesicetal2011} developed a \textit{graph literacy} scale composed of items that measured the ability to understand information presented in data viz in health-related contexts. The three stages model used to design the test was that by~\citet{friel2001making}, and was composed of the levels of  \textit{reading data}, \textit{interpreting relationships between data}, and \textit{extracting information beyond the data}.


\subsection{RQ3. Applied Research Reports}
\label{sec:applied}

We can identify trends across the nine reports that have been categorized as applied research; these trends include the focus on the user, citizen science, interpretation, and the use of DVL in various settings. However, there are differences in the areas of focus, the methods used, and the samples investigated. It is possible to group these nine reports based on the aim of the study.

\subsubsection{Understanding Users Behaviour}

Three reports focused on user behaviour in the context of data visualization. ~\citet{yang2021explaining} discussed how to teach data viz with an emphasis on introductions. ~\citet{rodrigues2021questions} investigated the mental processes of novices in analysing data viz by identifying their patterns and problems.  ~\citet{doi:10.1080/15391523.2018.1564636} focused on how digitally skilled students can engage in the construction of interactive data viz and the students' ability in data interpretation. 

\subsubsection{Citizen Science}

The second group of reports highlighted the need for contextually relevant strategies in enhancing DVL. ~\citet{peppler2021cultivating} examined how museums can foster DVL, noting the possibilities of learning outside of formal contexts. In the same manner,~\citet{borner2016investigating} explored how visitors of science museums engaged with data viz and the difficulties and possibilities of improving DVL in public spaces.

\subsubsection{Interpretation}

The third group of reports is related to data viz interpretation in various settings. ~\citet{binali2022high} investigated the correlation between students' educational level and their interpretation ability. ~\citet{aoyama2003graph} investigated the students’ ability to create new information from qualitative and quantitative information. These two reports support the identification of the factors that affect data viz comprehension and interpretation.

\subsubsection{Methodologies for Assessing DVL}
The fourth and last group of the applied research reports focused on assessing DVL. ~\citet{boden2019emerging} investigated the aspects of visual literacy that may be developed through engagements between visual analytics and students in social science classes; they focused on socio-material relations as the key to enhancing visual literacy. Similarly,~\citet{camba2022identifying} pointed out the importance of deception detection in DVL; they provided insights into the complexities of understanding data viz correctly. These two reports helped to advance the methodological development of assessment approaches to DVL by considering the assessment methods and their consequences.


\subsection{RQ4. Data Visualization Literacy Assessment}
\label{sec:assess}

In this section, we analyse the reports that proposed or used a questionnaire, or test, to measure DVL, taking into account the characteristics of the items (questions), the type of data viz considered, the sample size and the methods, if any, used to verify the reliability and validity of the proposed scales.

Among the 19 reports under analysis, fourteen~\citep{aoyama2003graph,Galesicetal2011,6875906,doi:10.1128/jmbe.v15i2.703,maltese2015data,garcia2016measuring,Lee2017551,krejci2020visual,locoro2021visual,yang2021explaining,camba2022identifying,firat2022p,reddy2023visual,davis2024risks} proposed an original test. Among the five remaining reports, four~\citep{lee2019correlation,binali2022high,Pandey20231,cui2023adaptive} mainly used modified versions of the VLAT test~\cite{Lee2017551}, which appears so far as the most frequently used test, whereas~\citet{Okanetal2019} proposed a short version of the~\citet{Galesicetal2011} scale.

The tests proposed by~\citet{reddy2023visual} and~\citet{garcia2016measuring} differ from the others because both are self-reporting tests.~\citet{ garcia2016measuring} designed a Subjective Graph Literacy (SGL) scale composed of 10 items (an example of which is: ``How good are you at working with bar charts?''), on a 6 point Likert scale, with 1 meaning ``not at all good'', and 6 ``extremely good''. In order to validate the scale, the SGL as well as the Objective Graph Literacy (OGL) tests~\citep{ Galesicetal2011} were administered to 400 subjects, recruited from MTurk. Garcia-Retamero and colleagues found that the SGL scale and its reduced version, composed of the first 5 items, were ``psychometrically efficient instruments that assess skill-related beliefs'', with raw scores moderately correlated with those from OGL.

Apart from the two self-reporting tests and~\citet{doi:10.1128/jmbe.v15i2.703}'s test, which was composed of a mix of images and data viz (line and bar charts), the questionnaires used in the remaining reports were composed of items referring to a specific data viz for which a task was requested and a response format was given. 

%
\subsubsection{Data Viz Used in Tests to Measure DVL}
\begin{figure*}[ht]
\centering
\includegraphics[width=0.8\textwidth]{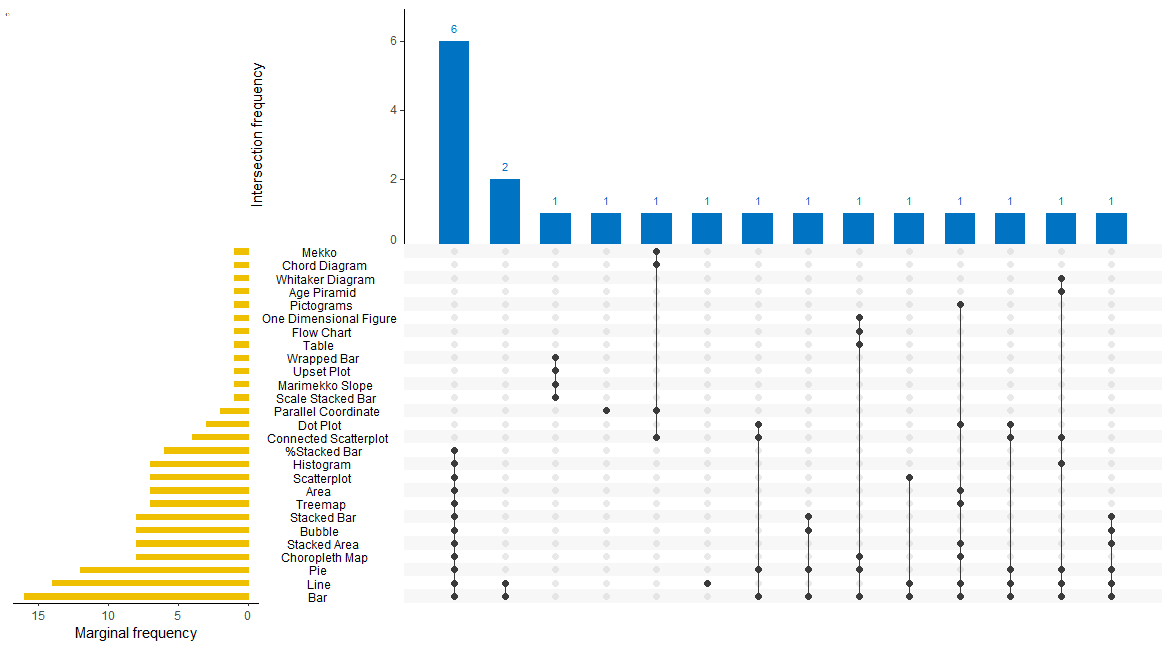}
\caption{Frequency distribution of data viz (yellow bars) and combination of data viz and tests (blue bars)}
\label{fig:Upset}
\end{figure*}

Questionnaires with data viz were present in 17 reports, and included a variety of data viz; Figure~\ref{fig:Upset} shows the frequency distribution of the data viz (yellow bars) and the frequency distribution of the combination of data viz and tests (blue bars).
Bar charts were the most common data viz; in one report, it was the only one used to evaluate DVL. Line and pie charts were the second most frequently used data viz. Most of the tests included easy data viz, which are, after all, the most common in scientific and daily contexts. The combination of data viz with higher frequency characterizes the VLAT test~\citep{Lee2017551}; however, different combinations of them, and sometimes the inclusion of more difficult data viz were explored by singular reports.

%
\subsubsection{The Task Required}
Related to each data viz in a test, a task was usually required to participants; it mirrors aspects of the measured DVL.~\citet{6875906} focused on six tasks, common to each data viz, which required ``purely visual operations or mental projections on a graphical representation'', i.e., to identify a maximum, a minimum, a variation (e.g detect trends, similarities or discrepancies), an intersection, an average (e.g. estimate an average point), and to make a comparison (e.g. between different trends or values). Instead,~\citet{Lee2017551} introduced a set of seven tasks, some of them in common with ~\citet{6875906}, which were data viz specific; apart from ``Find Extremum'', which was common to all of the data viz, different data viz had a different number of tasks.
The tasks considered were: retrieve a value, find an extremum, determine a range, find anomalies, clusters, correlations or trend and make comparisons. These tasks were also used by~\cite{yang2021explaining} in preparing their test.~\citet{aoyama2003graph},~\citet{Galesicetal2011} and~\citet{krejci2020visual} did not propose a set of tasks, as described above, but followed the~\citet{Frieletal2001} approach in writing the items. 
They grouped the items based on three comprehension levels: (1) the elementary level, which corresponds to the ability to read the data, namely to extract information from the data viz to answer to a question with obvious answer in the data viz, (2) the intermediate level, which corresponds to the ability to read between the data, namely to interpolate and find relationships in the data as shown on the data viz, and (3) the advanced level, which corresponds to the ability to read beyond the data, namely to extrapolate and analyse  relationships which are implicit in the data viz.
Even if they did not specify any task, looking at the text of the 6 items of the Aoyama and Stephens' scale, the 13 items of the Galesic and Garcia-Retamero's scale, and the 18 items of the Kreici et al.'s scale it was possible to recognize some of the tasks suggested by~\citet{6875906} and~\citet{Lee2017551}.
~\citet{locoro2021visual} proposed around 90 items based on evolutionary skills, partly inspired by the goal descriptions of the US scholar curricula of K8 educational path, and mapped with data viz of varied difficulty. These skills correspond to the levels of a construct map, described in Section~\ref{sec:basic}.

%
\subsubsection{Response Options}
The response options of the items were mainly multiple-choice, with only one true option and all the remaining ones as false, or of the binary (true-false) kind. For both the multiple-choice and the true-false items the responses were scored as 1 if correct and 0 if incorrect;~\citet{ maltese2015data} scored as 0 the missing answers too.~\citet{ Galesicetal2011} included items in which the test takers had to fill in the blank with a specific value, read from the data viz; the right answer was coded as 1 and, being it unquestionable, an agreement by multiple experts was not necessary. Two reports,~\citet{binali2022high} and~\citet{locoro2021visual}, included few constructed-responses items, e.g. items which required, as a response, a short free text. When a constructed-responses item was present in the test, its answers were examined and coded by some experts and an inter-rater reliability was measured, for example, by Cohen’s kappa.~\citet{Lee2017551} included in every item of VLAT the response option ``Omit'', which was provided, as they wrote, ``in order to address the issue of guessing in multiple-choice items''. Another strategy to protect from guessing was the introduction of the ``Don’t know'' answer, as done by~\citet{doi:10.1128/jmbe.v15i2.703}, who decided to treat this response as an incorrect response. 

\subsubsection{Tricking Items}
An ability dimension taken into account in the design of some tests devoted to measure DVL was that of recognizing and read beyond a misleading visualization, i.e., a visualization that could lead to conclusions not supported by the data.~\citet{Galesicetal2011},~\citet{camba2022identifying} and~\citet{cui2023adaptive} faced this issue, including in their test ``trick'' items, i.e., items whose data viz contained a deceptive visualization, in addition to normal items, namely items based on well-formed data viz. 
~\citet{Galesicetal2011} inserted the Q10 item, for which two bar charts, representing the same data but with different range scales, were requested to be compared, and the Q11 item, for which two line charts with unlabelled axis were requested to be compared. They found that these two items were the items with the lower percentage of correct responses.
~\citet{camba2022identifying} limited their analysis to consider the truncation or distortion of the Y-axis range as a deceptive visualization in a line chart, and proposed a test made up of half of the items related to deceptive data viz.
~\citet{cui2023adaptive} proposed the adaptive version of the CALVI test, composed by 11 ``trick'' and 4 normal items. CALVI was designed by~\citet{CALVI} to measure ``the ability to read, interpret, and reason about erroneous or potentially misleading data viz''. They identified eleven possible ``misleaders''  to intersect with nine chart types, i.e., four types of manipulation of scales (inappropriate order, inappropriate scale range, inappropriate use of scale functions and unconventional scale directions), inappropriate aggregation, misleading annotations, missing data, missing normalization, overplotting, concealed uncertainty and cherry picking (i.e. select a subset of data to display, which could be misleading when inference about the whole data set was required).~\citet{CALVI} properly highlighted that, whether a graph is misleading or not, depends on the task required from that graph, and not only on the misleading graph itself.

\subsubsection{Number of Items in a test}
\begin{figure}[ht]
\centering
\includegraphics[width=0.485\textwidth]{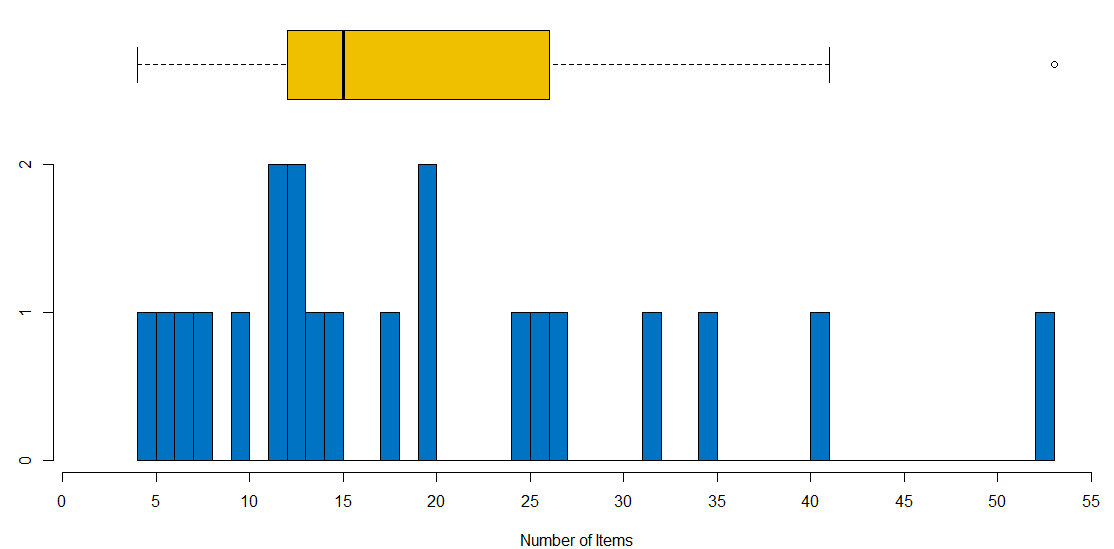}
\caption{Distribution of the reports with respect to the number of items involved in the questionnaire}
\label{fig:BoxPlot:items}
\end{figure}

Regarding the length of a test, Figure~\ref{fig:BoxPlot:items} reports the related frequency distribution based on 18 reports, i.e., all the reports except the~\citet{davis2024risks} test. The reason of this exclusion was related to the type of experiment that Davis and his coauthors performed. They selected 4 data viz types and 20 datasets, obtaining 80 items, then for each data viz-dataset pair, 15 repetitions were created, so 1200 items (trials in the paper) were administered to a sample of MTurk test takers. Clearly, a test made up of 1200 items is useless, if used as a tool to measure DVL, so it was not informative with respect of the aims of the present analysis.

The number of items forming the test varied from a minimum of 4 to a maximum of 53, with a median of 15 items; the longest questionnaire was the VLAT test~\cite{Lee2017551}. There is a trade-off between the test length and the quality of the final data; long tests could induce degradation in the responses due to the fatigue of the test takers; in this direction, the~\citet{Pandey20231} and~\citet{cui2023adaptive} works proposed some suggestions. The former proposed a shorter version of VLAT composed of 12 items, whereas the latter proposed an adaptive version of VLAT and CALVI.

\subsubsection{Content validity}
When designing a test to measure a latent trait, such as the DVL, it is necessary to assess the content validity of that test, i.e., evaluate the extent to which the test measures the latent trait of interest. In literature, there were mainly two approaches to trying out the reliability and validity of a test, Classical Test Theory (CTT) and Item Response Theory (IRT), with their specific instruments~\cite{Cappellerietal2014}. 
The majority of the reports (63.2\%) reported at least a reliability index and, among them, only two reports~\cite{6875906,cui2023adaptive} used an IRT model, whereas the others used Cronbach's $\alpha$ or McDonalds's $\omega$~\cite{Zinbargetal2005} alone, as well as with other tools of the CTT.

\subsubsection{Sample size}
\begin{figure}[ht]
\centering
\includegraphics[width=0.485\textwidth]{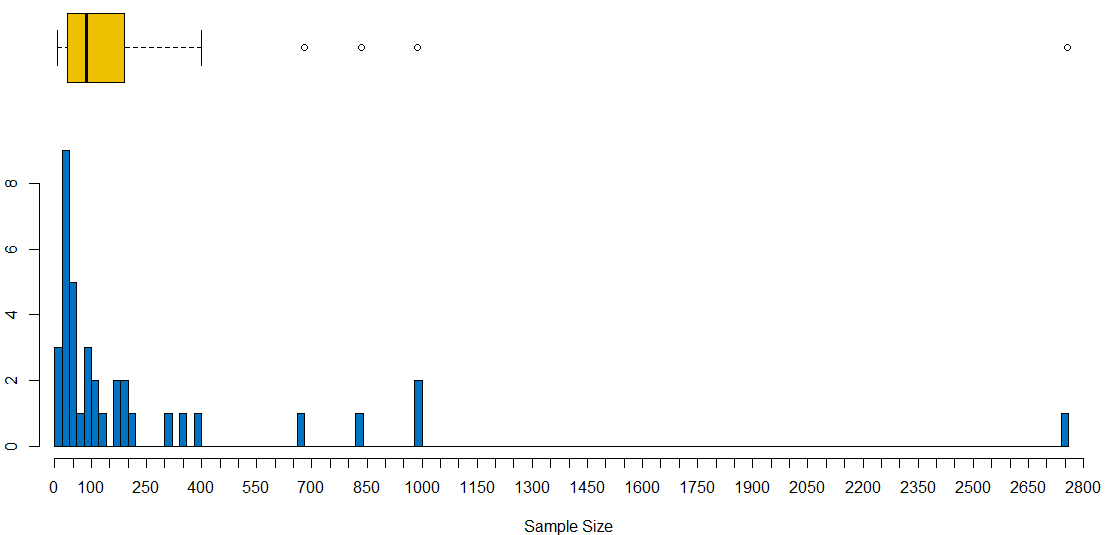}
\caption{Distribution of the experiments with respect to their sample size}
\label{fig:BoxPlot}
\end{figure}

This dimension is related to instrument validation: as far as the sample size increased, the results became more robust and stable. The 19 reports analyzed in this section included, in half cases, more than one experiment with dedicated samples. Figure~\ref{fig:BoxPlot} reports the sample size distribution of these experiments.

The distribution was highly and positively skewed, with a median of 86 test takers. The test takers were recruited via crowdsourcing platforms in the 53\% of these experiments, and between them six used MTurk and four Prolific.

\section{Discussion}
\label{sec:discussion}
As shown in Section~\ref{sec:results}, a theoretically fragmented landscape dominates the field of characterizing and measuring DVL. 
The research field of Data Visualization, whose aim was to design meaningful statistical patterns of the data, was the more prolific in the direction of setting the terrain for assessing the literacy of individuals for data viz. However, as highlighted above, in the reports analyzed neither there was an explicit characterization of the visualization ability nor of the data viz difficulty. Considering DVL as a scientific endeavour means affirming its meaning, trying to characterize it unambiguously, and modelling a construct that can be validated with a proper assessment procedure. For these reasons, we derive the following issues and research directions from the reports analyzed.

DVL was largely discussed without explicitly considering its characterization (i.e., the question ``What is Data Visualization Literacy?'' was never answered beyond the mere definition of the term). A characterization of the term is intended first of all rigorously. It regards the formal and ``ontological'' foundations of the DVL property, its necessary and sufficient determinants, and whether and how it is possible to measure them. The problem is not trivial, as having an explicit model of ``what people have to know and do'' to possess a good (or bad) level of DVL is far from being an easy conceptualization task. Many factors contribute to its complexity. Above all, the activity of measuring subjects, as their individuality, included their cultural background, past experiences and ever-changing skills during lifetime are at least moving targets. Furthermore, the same notion of measurement is questionable where it concerns human beings. On the other hand, measuring something without characterizing it explicitly may generate contradictions and does not help advance knowledge. For these reasons, we deemed crucial to make explicit any definition, model and critically consider any opaque approach currently available in the literature.


In the rapidly evolving landscape of 21st-century education, the integration of systematic and multidisciplinary visual literacy models is increasingly significant~\cite{vermeersch2015kids}. Deciding which visual literacy skills to teach can be a daunting task given the breadth of necessary competencies, from the identification of the need for an image, to its numerical meaning, and its ethical and legal understanding. This decision becomes even more complex when considering the inclusion of natural visuals and those embedded within media and communication, as both realms often use symbols and texts that are also central to data viz. The benefits of embedding DVL into educational frameworks were witnessed by studies proposing DVL interventions in STEM courses, to enhance critical thinking skills, and to improve students' understanding of complex concepts like global sustainability~\cite{mnguni2016assessment,arneson2018visual,doi:10.1128/jmbe.v15i2.703}. These interventions allowed students to analyze and interpret visual information more effectively, manipulate data into figures, and apply domain-specific knowledge crucial for informed decision-making. Although the adoption of DVL literacy across educational curricula demonstrated to lead to tangible benefits, strategies to connect images to course content, incorporate technology effectively, and potentially include participative instructor tools to deepen students' competencies are still rare. Instructors should foster this by using various visual aids, offering interpretive guidance, integrating real-world examples, facilitating group discussions, and providing feedback on students' DVL skills. However, the challenge extends beyond simply teaching these skills. In the broader context of DVL, there was a pressing need for accurate and reliable methods to assess individuals' capabilities. This necessity was underscored by researches designing or using conceptual models to pinpoint cognitive abilities essential for effective visualization. These models helped identify key cognitive DVL skills and devise appropriate instruments for assessing each pertinent visualization skill (e.g., ~\cite{mnguni2016assessment,locoro2021visual,Borner20191857}). Furthermore, many frameworks~\cite{Kedra201867,martin2022framework,doi:10.1128/jmbe.v15i2.703, marzal2020taxonomic,Borner20191857} were proposed to coordinate data viz and multiliteracy skills. These frameworks aimed to define the relationships between competencies necessary for educational innovation and alignment with academic curricula. They advocated for collaboration and integration of digital skills, supporting specialist training, and fostering a systematic approach to assessment tests. Interestingly, the field of DVL also acknowledged the need to go beyond the educational domain and consider individual differences in the visualization abilities of laypeople~\cite{bresciani2015pitfalls,lundgard2021accessible}, which can impact how data viz are perceived and interpreted. Recognizing these differences is crucial for creating personalized visualizations that cater to diverse needs and abilities, as suggested for example by~\cite{davis2024risks}. Practical classifications and models to enhance the accessibility and effectiveness of data viz were provided, ensuring they were comprehensible to a broad audience, including those with disabilities or limitations in visual perception. It is essential to ensure maximum clarity and correct interpretation of the information presented in graphical form, especially when this information may involve risks, such as in the communication of medical data to patients~\cite{Galesicetal2011}. This highlights the need to develop strategies and tools to improve people's DVL, for example in contexts related to health decisions. Furthermore, it was highlighted that understanding the information presented in graphical form requires a certain level of statistical thinking, a skill that may not be adequately taught~\cite{cui2021synergy}. Finally, it was deemed very important also to regulate the design of data viz. For example, the proposed classification by~\citet{bresciani2015pitfalls} has many practical implications: it can be used as a checklist to support designers in understanding overlooked issues in data viz, it can help users to check on their potential biases with the designer's work, and practitioners as an instructional tool for the evaluation, guidance for improvement, standard terminology, quality assurance, skill development, and error prevention when interacting with data viz.



Our analysis of the main results and trends observed on nine applied research reports on DVL reveals several key findings and trends.One of the most relevant findings was the importance of a user-centric focus; all nine reports have emphasised user-centric approaches. According to the nine reports analyzed, it is important to investigate and understand how users interpret visuals. Specifically, to improve DVL ~\citet{yang2021explaining, rodrigues2021questions} suggested the use of effective communication and specific instructional approaches. Another relevant finding is contextualisation. For instance, ~\citep{borner2016investigating, peppler2021cultivating} supports the need for context-specific DVL development. This result suggests that DVL education can be delivered beyond the classroom setting. A final finding is associated with the assessment of critical thinking. ~\citet{camba2022identifying} highlighted the importance of critical evaluation and deception detection in understanding visuals. This result suggests the need for assessment methods beyond simple knowledge checks; i.e.,  assessment methods shall include a more comprhenesive and complex analysis of DVL skills.

In the DVL context, the administration of a test was used to gain data to be used to measure the level of DVL owned by the test takers.
To do that, the majority of the 19 reports used the sum score (or raw score), i.e., the total count of the right responses given to the test items, directly or implicitly referring to what is done in CTT to measure the ``true score''. Nevertheless, as underlined by~\citet{BORSBOOM2002}, CTT does not assume that there is a construct (latent trait) underlying the measure, so the identification of the true score with the construct score is not, in general, compatible with CTT. The sum score has a second drawback; it is an ordinal variable, i.e. measured on an ordinal scale, so, as pointed out by~\citet{Merbitzetal1989} and~\citet{Grimbyetal2012}, the conclusions based on the results obtained from mathematical manipulations of the sum score may be misleading.
Measures of the latent trait based on the application of IRT models, such as, as an example, the Rasch or the 2-parameter logistic models~\cite{DeMars2010}, overcome these limits.
IRT models require a higher sample size than CTT, and this is one of the reasons why~\citet{Lee2017551} preferred CCT to IRT. Nevertheless, a high sample size is needed whenever a new questionnaire is proposed, so that the results of the validation step are more robust. From the analysis of the data viz used in the questionnaires, we found that bar, line e pie charts were the most used. These are data viz that are usually taught at school, not so difficult to interpret, given that at some point, everyone has met them, and highly common in scientific and daily contexts. A paradox was observed regarding the test length. The longer the test the higher the scale reliability. Nevertheless, a questionnaire with a high number of items could produce data of lower quality than a short test, due to the impact of a long test on the attention and accuracy of the test takers in responding to the questions. This aspect was considered by~\citet{Pandey20231} and~\citet{cui2023adaptive}, who proposed shorter versions of VLAT, which is composed of 53 items.

\subsection{Research directions}

A research direction emerging from our review is that of mitigating the proliferation of terms and the semantic heterogeneity of definitions, towards a (standard) unifying construct of DVL. The following factors could be considered:
\begin{itemize}
\item a standard nomenclature; 
\item a clear scope of what are the objects, dimensions and objectives at stake, together with the need to identify once and for all how to call this property; 
\item what data viz are included and what are excluded from DVL; 
\item how this property can be modelled according to evolving levels or measurable degrees. 
\end{itemize}

Furthermore, many satellite aspects, such as: 
\begin{itemize}
\item the evolving meaning of visual literacy and the role of technology, culture, and communication in this evolution; 
\item the usability of data viz and the role of design; 
\item the limitations of users and how to support them; 
\item the features and functionalities of the supporting technology to interact with data viz in an effective and efficient way 
\end{itemize}
should not be overlooked as fundamental preconditions of any attempt to characterize a construct empirically grounded, and not isolated from the rest.

Regarding the impact on researchers, our findings highlight the effectiveness of user-centered approaches, the need for diverse educational strategies, the importance of context-specific interventions, and the emphasis on developing assessment methods that go beyond basic comprehension and encourage critical thinking skills. According to these findings, future studies could investigate:
\begin{itemize}
\item how to customise DVL interventions to specific user profiles, considering variables like age, prior knowledge, and learning styles,
\item the long-term impact of different educational strategies (e.g., interactive creation vs. traditional instruction) on DVL skill retention and application,
\item	how to design and implement practical DVL learning experiences within museums and similar informal settings,
\item	methods that go beyond basic comprehension and assess a user's ability to critically analyse visuals, identify deception, and draw evidence-based conclusions.
\end{itemize}


Regarding the assessment of DVL, what was clearly emerging from our analysis are:
\begin{itemize}
\item the need to widen the spectrum of data viz that should be taken into consideration during experiments design for assessing DVL;
\item the need to improve the quality of experiments by the identification of the sample size, the improvement of items design, the application of more sophisticated statistical validation.
\end{itemize}

Besides bar charts, pie and line charts, many other charts could be considered as common (e.g., word clouds) and be included in tests, and new experiments could be started to assess whether and how data viz could be classified based on their difficulty level of encoding, decoding, and interpretation, and, based on this difficulty, their presence could be equally distributed in the design of new assessment tests. 

Regarding the reliability of tests, one possible research direction could aim at improving it without growing the number of items. For example, by considering the possibility of designing new tests where also polytomous items could be included, e.g., multiple choice items could be better designed, by changing their structure from one dichotomous style (one correct response as scored 1 / all the other incorrect responses as scored 0) into one polytomous style, with one correct response (scored 2), one partially correct response (scored 1), and one or more incorrect responses (scored 0).~\citet{Vispoel2014} have shown that the reliability of a test composed of the same number of items, but different type of responses, i.e., dichotomous vs polytomous (e.g. Likert scale) responses, is higher if the test is composed of polytomous items. 

\subsection{Limitations of the Study}
\label{sec:limitation}

The review is based on the literature of the last ten years. The decision to start from 2014 and not before was due to some important reasons. First of all, the review by~\citet{firat2022interactive} clearly defined that only preliminary studies existed before 2016, and mentioned the keynote speech of B{\"o}rner about Data Visualization Literacy in the venue of IEEE Vis 2019 as a starter of initiatives for assessing DVL from that moment on. Secondly, we considered that technological evolution either in the field of educational tools and in the field of Data Visualization design did not occur before 2014, if we consider that one of the the first tools for the professional development of data viz and dashboards were fully available since 2019 (e.g., Tableau\footnote{\url{https://www.tableau.com/}}). The two reports~\cite{aoyama2003graph,Galesicetal2011} that dates back to 2003 and 2011, respectively, included in our review, exemplify the level of infancy of both basic and applied research about DVL before the last decade. 

Another limitation may regard the fact that the terminology used to identify DVL in the literature is disparate, with many different terms denoting quite the same object of inquiry (see the search query used to retrieve the literature in Section~\ref{sec:extraction}). 
We avoided searching explicitly into the ``Keywords'' metadata where possible, as including it brought to many records (around 1,000) where the relevant terms (mostly ``Visual Literacy'') were only in the keywords, and a manual inspection of all of the title and abstract of those records revealed that they were all off topic. 
Furthermore, avoiding the use of many combinations of the term ``data viz'' with the word ``literacy'' (as including them all brought a too huge amount of reports to be manageable and tractable), may have brought to limitations in providing results in a fully complete search space. However, the direct experience of authors and experts consulted in the field, as well as the snowballing activity made on the retrieved reports, are an assurance that the most relevant research in the field were included in our review.

During the preliminary phases of our work, we considered including in the search either the reference conference on Data Visualization, IEEE VIS, or the ACM CHI conference on Human Factors in Computing Systems, the ones with the highest conference rank evaluation (i.e., A/A*, according to the Conference Ranking portal\footnote{available at \url{http://www.conferenceranks.com/}} and the Core Computing Research and Education Portal\footnote{available at  \url{https://www.core.edu.au/icore-portal}}). However, we soon realized that papers from the IEEE VIS conference were mostly published in the ``IEEE Transactions on Visualization and Computer Graphics'' Journal, after having been accepted for the conference. Regarding the CHI conference, we have launched the same query in the ACM portal, but no relevant results came from an inspection by two of the authors. Hence, we excluded conference records from our literature review. During the entire analysis of reports, either the manual addition of references or the snowballing procedure did not bring to our attention any conference record, making a piece of further evidence that our initial choice was fair.

Another critical aspect lays in the lack of a precise (standard) definition of DVL and of its models / dimensions. Many satellite aspects were considered in our review: the usability of data viz; the technology exploited and its functionality; the evolution of the discipline; and the strand of biometric measurements (e.g., eye-tracking) as an alternative or complementary activity to study the behavioral patterns of experts and non-experts users. These reports and their keywords were not explicitly retrieved and their analysis in our study should not be considered as a complete overview of them. Again, our choice was due to a tradeoff between completeness and accuracy. Our study focused on the definition, characterization, and assessment of DVL and, given the disparate perspectives present in the literature, framing and interpreting each report in light of this tripartition was painstaking but, obviously, subjective and relying on the interpretation and effort synergy of the authors. To alleviate potential inaccuracies in the data extraction and classification, we adopted a rigorous and systematic strategy made of several calibration tests, where each reviewer applied independently inclusion and exclusion criteria and then the results were crosschecked among all of the authors, until a complete agreement was reached. For each author, all doubtful situations were always discussed with the others to reach an agreement. 


\section{Conclusions}
\label{sec:conc}
This paper presented a systematic review on the characterization and measurement of DVL. To this end, the most important journals in the field, according to Elsevier, Scopus, and IEEE Explore have been scrutinized, finding 43 reports that provided an answer to our overarching research question, and to four derived questions. The analysis of these reports allowed us to identify the purposes of DVL, its satellite aspects, the models proposed, and the assessments designed to evaluate the degree of DVL of people. Several gaps have been founded in the literature: the majority of the selected reports did not adopt a unified view on DVL nor a systematic approach to the topic. The findings emerged from the review highlighted open issues and research directions such as the need to characterize DVL, the need to improve assessment tests in terms of better sampling and more sophisticated items design and analyses, and the need to adopt education and design strategies to strengthen the knowledge about and the usability of data viz as crucial tools for interpreting and managing the challenges of our times.

\section*{Acknowledgments}
Research funded by the European Union – Next-Generation EU - PRIN 2022 D.D. 104 of 02-02-2022 - Project name: ``Characterizing and Measuring Visual Information Literacy'' ID 2022JJ3PA5.

{\appendices

}


 

\begin{IEEEbiography}[{\includegraphics[width=1in,height=1in,clip,keepaspectratio]{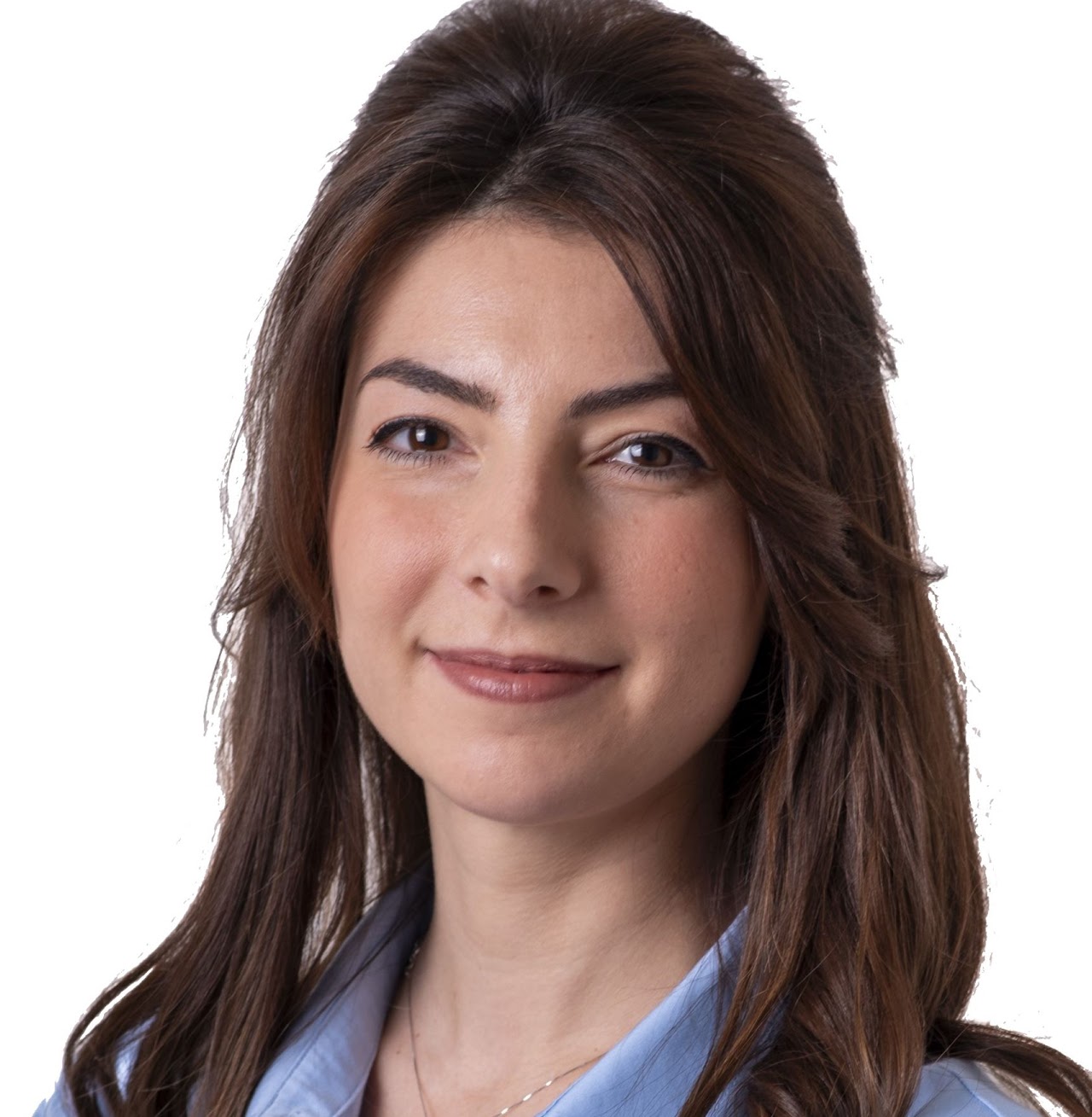}}]{Sara Beschi} is a Computer Science Engineer. She is an Associate Researcher at the University of Brescia, Italy, where she teaches Database Management. She is engaged in developing theories on Visual Information Literacy and Data Visualization. As a registered member of the Order of Engineers of Brescia, she brings extensive experience working on digital projects in a multinational company and a consulting agency specializing in data analytics, enhancing her contributions to these fields.\end{IEEEbiography}

\begin{IEEEbiography}[{\includegraphics[width=1in,height=1in,clip,keepaspectratio]{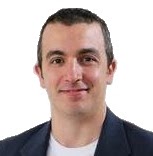}}]{Davide Falessi} is an Associate Professor of Software Engineering at the University of Rome Tor Vergata, Italy. He is the Associate Editor in Software Economics of IEEE Software and an Editorial Board member of the Empirical Software Engineering Journal.  His main research interest is in devising and empirically assessing scalable solutions for developing software-intensive systems. He received his Ph.D., MSc, and BSc degrees in Computer Engineering from the University of Rome Tor Vergata, Italy.\end{IEEEbiography}

\begin{IEEEbiography}[{\includegraphics[width=1in,height=1in,clip,keepaspectratio]{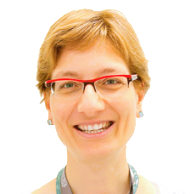}}]{Silvia Golia} is an Assistant Professor of Statistics at the University of Brescia, Italy. She received her MSc Degree in Statistics from the University of Padua and her PhD in Statistics from the University of Perugia. Her research interests include Rasch Analysis, Causal Analysis, Bayesian Networks, Decision Trees and Time series Analysis. \end{IEEEbiography}

\begin{IEEEbiography}[{\includegraphics[width=1in,height=1in,clip,keepaspectratio]{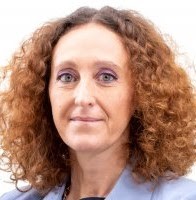}}]{Angela Locoro} is an Associate Professor of Artificial Intelligence at the University of Brescia, Italy. She is an Editorial Board Member of the Information Fusion Journal. Her main research interests are in the Data Visualization field, where she was funded for a project on Data Visualization Literacy. She received her Ph.D. in Computer Engineering, MA in Modern Literature, and BSc in Computer Science from the University of Genova. 
\end{IEEEbiography}


@article{page2021prisma,
  title={The PRISMA 2020 statement: an updated guideline for reporting systematic reviews},
  author={Page, Matthew J and McKenzie, Joanne E and Bossuyt, Patrick M and Boutron, Isabelle and Hoffmann, Tammy C and Mulrow, Cynthia D and Shamseer, Larissa and Tetzlaff, Jennifer M and Akl, Elie A and Brennan, Sue E and others},
  journal={Bmj},
  volume={372},
  year={2021},
  publisher={British Medical Journal Publishing Group}
}

@inproceedings{Ge2024Visualization,
  title={Toward a More Comprehensive Understanding of Visualization Literacy},
  author={Ge, Lily W. and Hedayati, Maryam and Cui, Yuan and Ding, Yiren and Bonilla, Karen and Joshi, Alark and Ottley, Alvitta and Bach, Benjamin and Kwon, Bum Chul and Rapp, David N. and others},
  booktitle={Proceedings of the CHI Conference on Human Factors in Computing Systems},
  year={2024},
  pages={7},
  organization={ACM},
  address={New York, NY, USA},
  doi={10.1145/3613905.3636289},
  month={May},
  venue={CHI EA '24},
  location={Honolulu, HI, USA}
}

@article{stevens1946theory,
  title={On the theory of scales of measurement},
  author={Stevens, Stanley Smith},
  journal={Science},
  volume={103},
  number={2684},
  pages={677--680},
  year={1946},
  publisher={American Association for the Advancement of Science}
}

@article{bertin1983semiology,
  title={Semiology of graphics},
  author={Bertin, Jacques},
  journal={Madison, WI: The University of Wisconsin Press.(Original work published 1967)},
  year={1983}
}

@book{tufte1990envisioning,
  title={Envisioning information},
  author={Tufte, Edward R and Goeler, Nora Hillman and Benson, Richard},
  volume={126},
  year={1990},
  publisher={Graphics press Cheshire, CT}
}

@book{mari2023measurement,
  title={Measurement across the sciences: Developing a shared concept system for measurement},
  author={Mari, Luca and Wilson, Mark and Maul, Andrew},
  year={2023},
  publisher={Springer Nature}
}

@article{cleveland1993model,
  title={A model for studying display methods of statistical graphics},
  author={Cleveland, William S},
  journal={Journal of Computational and Graphical Statistics},
  volume={2},
  number={4},
  pages={323--343},
  year={1993},
  publisher={Taylor \& Francis}
}

@ARTICLE{6875906,
  author={Boy, Jeremy and Rensink, Ronald A. and Bertini, Enrico and Fekete, Jean-Daniel},
  journal={IEEE Transactions on Visualization and Computer Graphics}, 
  title={A Principled Way of Assessing Visualization Literacy}, 
  year={2014},
  volume={20},
  number={12},
  pages={1963-1972},
  doi={10.1109/TVCG.2014.2346984}}

@article{doi:10.1128/jmbe.v15i2.703,
author = {Brian J. Rybarczyk and Kristen L.W. Walton and Wendy Heck Grillo},
title = {The Development and Implementation of an Instrument to Assess Students’ Data Analysis Skills in Molecular Biology },
journal = {Journal of Microbiology \&amp; Biology Education},
volume = {15},
number = {2},
pages = {259-267},
year = {2014},
doi = {10.1128/jmbe.v15i2.703},
}

@ARTICLE{Pandey20231,
	author = {Pandey, Saugat and Ottley, Alvitta},
	title = {Mini-VLAT: A Short and Effective Measure of Visualization Literacy},
	year = {2023},
	journal = {Computer Graphics Forum},
	volume = {42},
	number = {3},
	pages = {1 – 11},
	doi = {10.1111/cgf.14809},
	type = {Article},
	publication_stage = {Final},
	source = {Scopus},
}

@ARTICLE{Lee2017551,
	author = {Lee, Sukwon and Kim, Sung-Hee and Kwon, Bum Chul},
	title = {VLAT: Development of a Visualization Literacy Assessment Test},
	year = {2017},
	journal = {IEEE Transactions on Visualization and Computer Graphics},
	volume = {23},
	number = {1},
	pages = {551 – 560},
	doi = {10.1109/TVCG.2016.2598920},
	type = {Article},
	publication_stage = {Final},
	source = {Scopus},
}

@ARTICLE{Kedra201867,
	author = {Kedra, Joanna},
	title = {What does it mean to be visually literate? Examination of visual literacy definitions in a context of higher education},
	year = {2018},
	journal = {Journal of Visual Literacy},
	volume = {37},
	number = {2},
	pages = {67 – 84},
	doi = {10.1080/1051144X.2018.1492234}
}

@ARTICLE{Borner20191857,
	author = {B{\"o}rner, Katy and Bueckle, Andreas and Ginda, Michael},
	title = {Data visualization literacy: Definitions, conceptual frameworks, exercises, and assessments},
	year = {2019},
	journal = {Proceedings of the National Academy of Sciences of the United States of America},
	volume = {116},
	number = {6},
	pages = {1857 – 1864},
	doi = {10.1073/pnas.1807180116}
}

@article{aoyama2003graph,
  title={Graph interpretation aspects of statistical literacy: A Japanese perspective},
  author={Aoyama, Kazuhiro and Stephens, Max},
  journal={Mathematics Education Research Journal},
  volume={15},
  number={3},
  pages={207--225},
  year={2003},
  publisher={Springer}
}

@article{locoro2021visual,
  title={Visual information literacy: Definition, construct modeling and assessment},
  author={Locoro, Angela and Fisher, William P and Mari, Luca},
  journal={IEEE access},
  volume={9},
  pages={71053--71071},
  year={2021},
  publisher={IEEE}
}

@article{rodrigues2021questions,
  title={What questions reveal about novices’ attempts to make sense of data visualizations: Patterns and misconceptions},
  author={Rodrigues, Ariane Moraes Bueno and Barbosa, Gabriel Diniz Junqueira and Lopes, H{\'e}lio C{\^o}rtes Vieira and Barbosa, Simone Diniz Junqueira},
  journal={Computers \& Graphics},
  volume={94},
  pages={32--42},
  year={2021},
  publisher={Elsevier}
}

@article{firat2022p,
  title={P-lite: A study of parallel coordinate plot literacy},
  author={Firat, Elif E and Denisova, Alena and Wilson, Max L and Laramee, Robert S},
  journal={Visual Informatics},
  volume={6},
  number={3},
  pages={81--99},
  year={2022},
  publisher={Elsevier}
}

@article{peppler2021cultivating,
  title={Cultivating data visualization literacy in museums},
  author={Peppler, Kylie and Keune, Anna and Han, Ariel},
  journal={Information and Learning Sciences},
  volume={122},
  number={1/2},
  pages={1--16},
  year={2021},
  publisher={Emerald Publishing Limited}
}

@article{camba2022identifying,
  title={Identifying deception as a critical component of visualization literacy},
  author={Camba, Jorge D and Company, Pedro and Byrd, Vetria L},
  journal={IEEE Computer Graphics and Applications},
  volume={42},
  number={1},
  pages={116--122},
  year={2022},
  publisher={IEEE}
}

@article{firat2022vislite,
  title={VisLitE: visualization literacy and evaluation},
  author={Firat, Elif and Joshi, Alark and Laramee, Robert},
  journal={IEEE Computer Graphics and Applications},
  volume={42},
  number={3},
  pages={99--107},
  year={2022},
  publisher={IEEE}
}

@article{borner2016investigating,
  title={Investigating aspects of data visualization literacy using 20 information visualizations and 273 science museum visitors},
  author={B{\"o}rner, Katy and Maltese, Adam and Balliet, Russell Nelson and Heimlich, Joe},
  journal={Information Visualization},
  volume={15},
  number={3},
  pages={198--213},
  year={2016},
  publisher={SAGE Publications Sage UK: London, England}
}

@article{krejci2020visual,
  title={Visual literacy intervention for improving undergraduate student critical thinking of global sustainability issues},
  author={Krejci, Sarah E and Ramroop-Butts, Shirma and Torres, Hector N and Isokpehi, Raphael D},
  journal={Sustainability},
  volume={12},
  number={23},
  year={2020},
  publisher={MDPI}
}

@article{binali2022high,
  title={High school and college students’ graph-interpretation competence in scientific and daily contexts of data visualization},
  author={Binali, Theerapong and Chang, Ching-Hwa and Chang, Yen-Jung and Chang, Hsin-Yi},
  journal={Science \& Education},
  pages={1--23},
  year={2022},
  publisher={Springer}
}

@article{arneson2018visual,
  title={Visual literacy in Bloom: Using Bloom’s taxonomy to support visual learning skills},
  author={Arneson, Jessie B and Offerdahl, Erika G},
  journal={CBE—Life Sciences Education},
  volume={17},
  number={1},
  pages={ar7},
  year={2018},
  publisher={Am Soc Cell Biol}
}

@article{ruf2023literature,
  title={A Literature Review Comparing Experts’ and Non-Experts’ Visual Processing of Graphs during Problem-Solving and Learning},
  author={Ruf, Verena and Horrer, Anna and Berndt, Markus and Hofer, Sarah Isabelle and Fischer, Frank and Fischer, Martin R and Zottmann, Jan M and Kuhn, Jochen and K{\"u}chemann, Stefan},
  journal={Education Sciences},
  volume={13},
  number={2},
  pages={216},
  year={2023},
  publisher={MDPI}
}

@inproceedings{kim2021accessible,
  title={Accessible visualization: Design space, opportunities, and challenges},
  author={Kim, Nam Wook and Joyner, Shakila Cherise and Riegelhuth, Amalia and Kim, Y},
  booktitle={Computer Graphics Forum},
  volume={40},
  number={3},
  pages={173--188},
  year={2021},
  organization={Wiley Online Library}
}

@article{ansari2022enhancing,
  title={Enhancing the usability and usefulness of open government data: A comprehensive review of the state of open government data visualization research},
  author={Ansari, Bahareh and Barati, Mehdi and Martin, Erika G},
  journal={Government Information Quarterly},
  volume={39},
  number={1},
  pages={101657},
  year={2022},
  publisher={Elsevier}
}

@article{vermeersch2015kids,
  title={Kids, take a look at this! Visual Literacy Skills in the School Curriculum},
  author={Vermeersch, Lode and Vandenbroucke, Anneloes},
  journal={Journal of Visual Literacy},
  volume={34},
  number={1},
  pages={106--130},
  year={2015},
  publisher={Taylor \& Francis}
}

@article{thompson2020uniting,
  title={Uniting the field: using the ACRL Visual Literacy Competency Standards to move beyond the definition problem of visual literacy},
  author={Thompson, Dana Statton and Beene, Stephanie},
  journal={Journal of Visual Literacy},
  volume={39},
  number={2},
  pages={73--89},
  year={2020},
  publisher={Taylor \& Francis}
}

@article{marzal2020taxonomic,
  title={A taxonomic proposal for multiliteracies and their competences},
  author={Garc{\'\i}a-Quismondo, Miguel {\'A}ngel Marzal},
  journal={Profesional de la informaci{\'o}n/Information Professional},
  volume={29},
  number={4},
  year={2020},
  publisher={EPI SCP}
}

@article{davis2024risks,
  title={The risks of ranking: Revisiting graphical perception to model individual differences in visualization performance},
  author={Davis, Russell and Pu, Xiaoying and Ding, Yiren and Hall, Brian D and Bonilla, Karen and Feng, Mi and Kay, Matthew and Harrison, Lane},
  journal={IEEE Transactions on Visualization and Computer Graphics},
  volume={30},
  number={3},
  pages={1756--1771},
  year={2024},
  publisher={IEEE}
}

@article{martin2022framework,
  title={A framework for visual literacy competences in engineering education},
  author={Mart{\'\i}n Erro, Alfonso and Nuere Men{\'e}ndez-Pidal, Silvia and D{\'\i}az-Obreg{\'o}n Cruzado, Ra{\'u}l and Acitores Suz, Adela},
  journal={Journal of Visual Literacy},
  volume={41},
  number={2},
  pages={132--152},
  year={2022},
  publisher={Taylor \& Francis}
}

@article{yang2021explaining,
  title={Explaining with examples: Lessons learned from crowdsourced introductory description of information visualizations},
  author={Yang, Leni and Xiong, Cindy and Wong, Jason K and Wu, Aoyu and Qu, Huamin},
  journal={IEEE Transactions on Visualization and Computer Graphics},
  volume={29},
  number={3},
  pages={1638--1650},
  year={2021},
  publisher={IEEE}
}

@article{bhat2023infographics,
  author={Bhat, Sameer Ahmad and Alyahya, Suzan},
  journal={IEEE Access}, 
  title={Infographics in Educational Settings: A Literature Review}, 
  year={2024},
  volume={12},
  number={},
  pages={1633-1649},
  doi={10.1109/ACCESS.2023.3348083}
}

@article{blummer2015some,
  title={Some visual literacy initiatives in academic institutions: A literature review from 1999 to the present},
  author={Blummer, Barbara},
  journal={Journal of Visual Literacy},
  volume={34},
  number={1},
  pages={1--34},
  year={2015},
  publisher={Taylor \& Francis}
}

@article{kalaf2023promoting,
  title={Promoting information literacy and visual literacy skills in undergraduate students using infographics},
  author={Kalaf-Hughes, Nicole},
  journal={PS: Political Science \& Politics},
  volume={56},
  number={2},
  pages={321--327},
  year={2023},
  publisher={Cambridge University Press}
}

@article{reddy2023visual,
  title={Visual literacy shown through a magnifying lens by high school students},
  author={Reddy, Pritika and Sharma, Bibhya and Chaudhary, Kaylash and Lolohea, Osaiasi and Tamath, Robert},
  journal={Interactive Technology and Smart Education},
  volume={20},
  number={4},
  pages={493--511},
  year={2023},
  publisher={Emerald Publishing Limited}
}

@article{bresciani2015pitfalls,
  title={The pitfalls of visual representations: A review and classification of common errors made while designing and interpreting visualizations},
  author={Bresciani, Sabrina and Eppler, Martin J},
  journal={Sage Open},
  volume={5},
  number={4},
  year={2015},
  publisher={SAGE Publications Sage CA: Los Angeles, CA}
}

@article{mnguni2016assessment,
  title={Assessment of visualisation skills in biochemistry students},
  author={Mnguni, Lindelani and Sch{\"o}nborn, Konrad and Anderson, Trevor},
  journal={South African Journal of Science},
  volume={112},
  number={9-10},
  pages={1--8},
  year={2016},
  publisher={Academy of Science of South Africa}
}

@article{lee2019correlation,
  title={The correlation between users’ cognitive characteristics and visualization literacy},
  author={Lee, Sukwon and Kwon, Bum Chul and Yang, Jiming and Lee, Byung Cheol and Kim, Sung-Hee},
  journal={Applied Sciences},
  volume={9},
  number={3},
  pages={488},
  year={2019},
  publisher={MDPI}
}

@article{firat2022interactive,
  title={Interactive visualization literacy: The state-of-the-art},
  author={Firat, Elif E and Joshi, Alark and Laramee, Robert S},
  journal={Information Visualization},
  volume={21},
  number={3},
  pages={285--310},
  year={2022},
  publisher={SAGE Publications Sage UK: London, England}
}

@article{cui2023adaptive,
  author={Cui, Yuan and Ge, Lily W. and Ding, Yiren and Yang, Fumeng and Harrison, Lane and Kay, Matthew},
  journal={IEEE Transactions on Visualization and Computer Graphics}, 
  title={Adaptive Assessment of Visualization Literacy}, 
  year={2024},
  volume={30},
  number={1},
  pages={628-637},
  doi={10.1109/TVCG.2023.3327165}}

@article{boden2019emerging,
  title={Emerging Visual Literacy through Enactments by Visual Analytics and Students.},
  author={Bod{\'e}n, Ulrika and Stenliden, Linn{\'e}a},
  journal={Designs for Learning},
  volume={11},
  number={1},
  pages={40--51},
  year={2019},
  publisher={ERIC}
}

@article{cui2021synergy,
  title={Synergy between research on ensemble perception, data visualization, and statistics education: A tutorial review},
  author={Cui, Lucy and Liu, Zili},
  journal={Attention, Perception, \& Psychophysics},
  volume={83},
  pages={1290--1311},
  year={2021},
  publisher={Springer}
}

@article{lundgard2021accessible,
  title={Accessible visualization via natural language descriptions: A four-level model of semantic content},
  author={Lundgard, Alan and Satyanarayan, Arvind},
  journal={IEEE transactions on visualization and computer graphics},
  volume={28},
  number={1},
  pages={1073--1083},
  year={2021},
  publisher={IEEE}
}

@article{doi:10.1080/15391523.2018.1564636,
author = {Linnéa Stenliden, Ulrika Bodén and Jörgen Nissen},
title = {Students as Producers of Interactive Data Visualizations—Digitally Skilled to Make Their Voices Heard},
journal = {Journal of Research on Technology in Education},
volume = {51},
number = {2},
pages = {101--117},
year = {2019},
publisher = {Routledge},
doi = {10.1080/15391523.2018.1564636}
}

@article{maltese2015data,
  title={Data visualization literacy: Investigating data interpretation along the novice—expert continuum},
  author={Maltese, Adam V and Harsh, Joseph A and Svetina, Dubravka},
  journal={Journal of College Science Teaching},
  volume={45},
  number={1},
  pages={84--90},
  year={2015},
  publisher={JSTOR}
}

@article{garcia2016measuring,
  title={Measuring graph literacy without a test: A brief subjective assessment},
  author={Garcia-Retamero, Rocio and Cokely, Edward T and Ghazal, Saima and Joeris, Alexander},
  journal={Medical Decision Making},
  volume={36},
  number={7},
  pages={854--867},
  year={2016},
  publisher={Sage Publications Sage CA: Los Angeles, CA}
}

@article{Galesicetal2011,
author = {Mirta Galesic and Rocio Garcia-Retamero},
title ={Graph Literacy: A Cross-Cultural Comparison},
journal = {Medical Decision Making},
volume = {31},
number = {3},
pages = {444-457},
year = {2011},
doi = {10.1177/0272989X10373805},
 }

@article{Okanetal2019,
author = {Yasmina Okan and Eva Janssen and Mirta Galesic and Erika A. Waters},
title ={Using the Short Graph Literacy Scale to Predict Precursors of Health Behavior Change},
journal = {Medical Decision Making},
volume = {39},
number = {3},
pages = {183-195},
year = {2019},
doi = {10.1177/0272989X19829728},
}

@article{Frieletal2001,
author = {Susan N. Friel and Frances R. Curcio and George W. Bright},
title ={Making Sense of Graphs: Critical Factors Influencing Comprehension and Instructional Implications},
journal = {Journal for Research in Mathematics Education},
volume = {32},
number = {2},
pages = {124-158},
year = {2001},
doi = {doi.org/10.2307/749671},
}

@inproceedings{CALVI,
author = {Ge, Lily W. and Cui, Yuan and Kay, Matthew},
title = {CALVI: Critical Thinking Assessment for Literacy in Visualizations},
year = {2023},
isbn = {9781450394215},
publisher = {Association for Computing Machinery},
address = {New York, NY, USA},
doi = {10.1145/3544548.3581406},
booktitle = {Proceedings of the 2023 CHI Conference on Human Factors in Computing Systems},
articleno = {815},
numpages = {18},
location = {, Hamburg, Germany, },
series = {CHI '23}
}

@article{Zinbargetal2005,
  title={Cronbach’s α, Revelle’s β, and Mcdonald’s ωH: their relations with each other and two alternative conceptualizations of reliability},
  author={Zinbarg, Richard E and Revelle, William and Yovel, Iftah and Li, Wen},
  journal={Psychometrika},
  volume={70},
  number={1},
  pages={123--133},
  year={2005},
}

@article{Merbitzetal1989,
  title={Ordinal scales and foundations of misinference},
  author={Merbitz, Charles and Morris, Jeri and Grip, Jeffrey C.},
  journal={Archives of Physical Medicine and Rehabilitation},
  volume={70},
  number={4},
  pages={308--312},
  year={1989},
}

@article{Grimbyetal2012,
  title={The use of raw scores from ordinal scales: time to end malpractice?},
  author={Grimby, Gunnar and Tennant, Alan and Tesio, Luigi},
  journal={Journal of Rehabilitation Medicine},
  volume={44},
  number={2},
  pages={97-98},
  year={2012},
}

@article{Cappellerietal2014,
title = {Overview of Classical Test Theory and Item Response Theory for the Quantitative Assessment of Items in Developing Patient-Reported Outcomes Measures},
journal = {Clinical Therapeutics},
volume = {36},
number = {5},
pages = {648-662},
year = {2014},
doi = {https://doi.org/10.1016/j.clinthera.2014.04.006},
author = {Joseph C. Cappelleri and J. {Jason Lundy} and Ron D. Hays},
}

@article{BORSBOOM2002,
title = {True scores, latent variables, and constructs: A comment on Schmidt and Hunter},
journal = {Intelligence},
volume = {30},
number = {6},
pages = {505-514},
year = {2002},
doi = {https://doi.org/10.1016/S0160-2896(02)00082-X},
author = {Denny Borsboom and Gideon J. Mellenbergh},
}

@article{Vispoel2014,
title = {Psychometric properties for the Balanced Inventory of Desirable Responding: Dichotomous versus polytomous conventional and IRT scoring},
journal = {Psychological Assessment},
volume = {26},
number = {3},
pages = {878--891},
year = {2014},
doi = {https://doi.org/10.1037/a0036430},
author = {Walter P. Vispoel and Han Yi Kim},
}

@book{DeMars2010,
  title={Item Response Theory},
  author={Christine DeMars},
  year={2010},
  publisher={Oxford University Press}
}

@article{friel2001making,
  title={Making sense of graphs: Critical factors influencing comprehension and instructional implications},
  author={Friel, Susan N and Curcio, Frances R and Bright, George W},
  journal={Journal for Research in mathematics Education},
  volume={32},
  number={2},
  pages={124--158},
  year={2001},
  publisher={National Council of Teachers of Mathematics}
}

@article{von2019theoretical,
  title={Theoretical Foundations of Design Thinking: Part II: Robert H. McKim’s Need-Based Design Theory},
  author={von Thienen, Julia PA and Clancey, William J and Meinel, Christoph},
  journal={Design Thinking Research: Looking Further: Design Thinking Beyond Solution-Fixation},
  pages={13--38},
  year={2019},
  publisher={Springer}

}
\end{document}